\documentclass{nature}
\usepackage[pdftex]{graphicx}
\graphicspath{{./}}

\makeatletter
\let\saved@includegraphics\includegraphics
\AtBeginDocument{\let\includegraphics\saved@includegraphics}
\renewenvironment*{figure}{\@float{figure}}{\end@float}
\makeatother

\usepackage{amsmath,amsfonts,amsthm,amssymb}
\usepackage{xcolor}
\usepackage[pdftex,colorlinks=true,linkcolor=blue,citecolor=blue,urlcolor=blue]{hyperref}
\usepackage[capitalise]{cleveref}
\usepackage[utf8]{inputenc}
\usepackage[english]{babel}
\addto\captionsenglish{}

\usepackage{ccaption}

\def\rmd{{\mathrm{d}}}
\def\rme{{\mathrm{e}}}

\newcommand{\kbT}{k_\textnormal{B} T}

\usepackage{pdfpages} 
\usepackage{pgffor} 

\makeatletter
\AtBeginDocument{\let\LS@rot\@undefined}
\makeatother

\def\supplementfilename{TF_supplement.pdf}

\pdfximage{\supplementfilename}
\def\numbersupplementpages{\the\pdflastximagepages}

\newif\ifarXiv
\arXivtrue

\title{Electronic screening using a virtual Thomas–Fermi fluid
    for predicting wetting and phase transitions of ionic liquids at metal
    surfaces
}

\author{Alexander Schlaich$^{1,2\,\bigstar}$, Dongliang Jin$^{1}$,
	Lyderic Bocquet$^{3}$ \& Benoit Coasne$^{1\,\bigstar}$}

\begin{document}

	\maketitle

	\begin{affiliations}
		\item{Univ.\ Grenoble Alpes, CNRS, LIPhy, 38000 Grenoble, France}
		\item{University of Stuttgart, Allmandring 3, 70569 Stuttgart, Germany}
		\item{Laboratoire de Physique de l’Ecole Normale Supérieure, CNRS,
			Université PSL, Sorbonne Université, Sorbonne Paris Cité, Paris,
			France}
		\item[$^\bigstar$] e-mail: schlaich@icp.uni-stuttgart.de and
		benoit.coasne@univ-grenoble-alpes.fr
	\end{affiliations}

	\begin{abstract}
        {Of relevance to energy storage, electrochemistry and catalysis,
        ionic/dipolar liquids display unexpected behaviors --- especially in
        confinement. Beyond adsorption, overscreening and crowding effects,
        experiments have highlighted novel phenomena such as unconventional
        screening and the impact of the electronic nature --- metallic versus
        insulating --- of the confining surface. Such behaviors, which
        challenge existing frameworks, highlight the need for tools to fully
        embrace the properties of confined liquids. Here, we introduce a
        novel approach describing electronic screening while capturing
        molecular aspects of interfacial fluids. While available strategies
        consider perfect metal or insulator surfaces, we build on the
        Thomas–Fermi formalism to develop an effective approach dealing with
        any imperfect metal between these asymptotes. Our approach describes
        electrostatic interactions within the metal through a `virtual'
        Thomas-Fermi fluid of charged particles, whose Debye length sets the
        screening length $\lambda$. We show that this method captures the
        electrostatic interaction decay and electrochemical behavior upon
        varying $\lambda$. By applying this strategy to an ionic liquid, we
        unveil a wetting transition upon switching from insulating to
        metallic conditions.}
	\end{abstract}

	\maketitle

	The fluid/solid interface as encountered in confined liquids is the
\textit{locus} of a broad spectrum of microscopic phenomena such as
molecular adsorption, chemical reaction, and interfacial
slippage.\cite{bocquet_2010_nanofluidics} These molecular mechanisms are
key to nanotechnologies where the fluid/solid interaction specificities are
harnessed for energy storage, catalysis, lubrication, depollution, etc.
From a fundamental viewpoint, the behavior of nanoconfined fluids often
challenges existing frameworks even when simple liquids are considered.
Ionic systems, either in their liquid or solid state, between charged or
neutral surfaces lead to additional ion adsorption, crowding/overscreening,
surface transition, and chemical
phenomena\cite{schoch_2008_transport,bazant_2011_double,smith_2016_electrostatic,laine_2020_nanotribology}
 that are crucial in
electrokinetics (e.g.\
electrowetting) and electrochemistry (e.g.\
supercapacitors/batteries).\cite{fedorov_2014_ionic}
Theoretical descriptions of nanoconfined fluids --- except rare
contributions\cite{kaiser_2017_electrostatic,dufils_2019_semiclassical,
    newns_1969_fermi,inkson_1973_manybody,kornyshev_1977_image,
    luque_2012_electric,kornyshev_2014_differential,netz_1999_debyehuckel}
--- assume either perfectly metallic or insulating confining surfaces
but these asymptotic limits do not fully reflect real materials as they
display an intermediate imperfect metal/insulator behavior (only few
metals behave  perfectly and all insulators are semi-conducting to some
extent). Yet, the electrostatic boundary condition imposed by the
surrounding medium strongly impacts confined dipolar and, even more,
charged systems.\cite{lee_2016_ion,
    bedrov_2019_molecular,breitsprecher_2015_electrode}
For instance, the confinement-induced shift in the freezing of an ionic
liquid was found to drastically depend on the surface metallic/insulating
nature.\cite{comtet_2017_nanoscale}

Formally, quantum effects leading to electronic screening in the confining
metallic
walls can be accounted for using the microscopic Thomas--Fermi (TF)
model.\cite{ashcroft_1976_solid,comtet_2017_nanoscale}
This formalism relies on a local density approximation for the charges in
the metal which are treated as a free electron gas (therefore, restricting
the electron energy to its kinetic contribution). This simple
model allows considering any real metal --- from perfect metal to insulator
--- through the Thomas--Fermi screening length $\lambda$.
The latter is defined in terms of the electronic density of state of the
metal at the Fermi level $\mathcal{D}(\mathcal{E}_\mathrm{F})$ according to
$\lambda = {\varepsilon}/{e^2 \mathcal{D}(\mathcal{E}_\mathrm{F})}$
($\varepsilon$ is the dielectric constant and $e$ the elementary charge);
the Fermi energy is directly related to the free electron density $n_0$ as
$\mathcal{E}_\mathrm{F} = \hbar^2(3\pi^2 n_0)^{2/3}/(2m_e)$
where $m_e$ is the electron mass and $\hbar=h/2\pi$ the Planck constant,
see \textit{Supplementary Information II A}.
Despite this available framework, the development of classical molecular
simulation methods to understand the microscopic behavior of classical
fluids in contact with imperfect metals is only nascent.
While insulators are treated using solid atoms with constant charge, metals
must be described using an effective screening approach. The charge image
concept can be used for perfectly metallic and planar
surfaces\cite{dossantos_2017_simulations} but refined
strategies must be implemented for non-planar surfaces such as
variational\cite{siepmann_1995_influence,reed_2007_electrochemical} or
Gauss law\cite{tyagi_2010_iterative,arnold_2013_efficient,
    nguyen_2019_incorporating}
approaches to model the induced charge distribution in the metal.
A recent proposal\cite{dufils_2019_semiclassical} builds on our TF
framework\cite{kaiser_2017_electrostatic} to propose a computational
approach based on variational localized surface charges that accounts for
electrostatic interactions close to imperfect metals.

Here, we develop an effective yet robust atom-scale simulation approach
which allows considering the confinement of dipolar or charged fluids
between metallic surfaces of any geometry and/or electronic screening
length. Following Torrie and Valleau's work for electrolyte
interfaces,\cite{torrie_1986_double} the electronic screening in the
imperfect confining metal is accounted for
through the response of a high temperature virtual Thomas--Fermi fluid made
up of light charged particles. Due to its very fast
response, this effective Thomas--Fermi fluid mimics metal induction within
the confining surfaces upon sampling the confined system configurations
using Monte Carlo or molecular dynamics simulations. After straightforward
implementation in existing simulation packages, this strategy provides a
mean to impose a Thomas--Fermi screening length that
is directly
linked to the equivalent virtual fluid Debye length. By adopting a
molecular level description, our approach captures all specificities
inherent to confined and vicinal fluids at the nanoscale (e.g. density
layering, slippage, non-viscous effects).
This virtual Thomas--Fermi model correctly captures electrostatic screening
within the confined system upon varying the confining host from perfect
metal to insulator conditions. The presented molecular approach is also shown
to accurately mimic the expected capacitive behavior, therefore opening
perspectives for the atom-scale simulation of electrochemical devices
involving metals of various screening lengths/geometries. To further assess
this effective strategy, by considering the freezing of a confined ionic
liquid, we show that the experimental behavior observed in
Ref.\cite{comtet_2017_nanoscale} can be rationalized as the salt is predicted
to melt at a temperature larger than the bulk melting point and that the
shift in the melting point should increase with decreasing the screening
length. Last but not least, using this novel method, by considering an
ionic liquid between two parallel planes, we unravel a continuous wetting
transition as the surfaces are tuned from insulating (non-wetting) to
metallic (wetting).

A few remarks are in order. Formally, the impact of image forces induced in
non-insulating solid surfaces when setting them in contact with charged
and/or dipolar molecules was investigated in depth by Kornyshev and
coworkers.\cite{kornyshev_1980_nonlocal,vorotyntsev_1986_modern}
Using a formal treatment of the Coulomb interactions with appropriate
electrostatic boundary conditions, these authors have considered complex
problems ranging from electric double
layers\cite{kornyshev_1980_electrostatic,vorotyntsev_1988_} to ion/molecule
adsorption.\cite{kornyshev_1986_coverage,vorotyntsev_1980_possible}
This robust framework was employed for simple point charges but also for
complex systems such as ionic liquids (whose chemical structure leads to
ion-specific effects beyond electrostatic interactions).
In contrast, the Thomas--Fermi model --- which is at the heart of our
molecular dynamics approach --- is based on a simplified description of
conductivity in solids. This simple formalism allows covering situations from
perfect and imperfect metals to semiconductors by varying the screening
length $\lambda$.
The Thomas--Fermi model is known to be a rather crude approximation when
describing the interaction between ions and solid
surfaces.\cite{kornyshev_1989_metal}
In particular, due to the use of semi-quantum description level of charges in
solids, it fails to accurately describe quantum effects such as Friedel
oscillations, the image plane position and motion with respect to metal
atoms, and the electron spillover outside the metal (\textit{a fortiori}, the
impact of charge adsorption at the solid surface is even more complicated to
account for). As a result, in some cases such as for transition metals,
metallic surfaces will not be appropriately described using the Thomas--Fermi
approximation.

Yet, despite these drawbacks, as shown here, the Thomas--Fermi model provides
a simple framework to implement the complex electrostatic interaction ---
including image forces --- arising in the vicinity of metallic surfaces. Such
multicontribution interaction, which results from a theoretical derivation
described below (and in more detail in the \textit{Supplementary Information
II}), includes time demanding numerical estimates that cannot be calculated
on the fly in molecular dynamics. As a result, despite its limitations and
weaknesses, the Thomas--Fermi approach is an approximate scheme to account
for electrostatic screening near solid surfaces with properties that range
from perfectly metallic to insulating. In fact, in the present molecular
approach, rather than formally implementing the equations corresponding to
the Thomas--Fermi model, we rely on a simple effective procedure in which
electrostatic screening induced by a fluid of charges located in the solid
material is mapped onto the Thomas--Fermi length. As shown in this paper, we
believe that this matching is a valid approach as the mapped screening length
is consistent with the observed capacitance behavior of the system.
Therefore, despite its simplicity, we believe that our effective molecular
strategy provides a simple tool to investigate complex surface electrostatic
phenomena occurring at solid/liquid interfaces.
Even if this approach provides a first order picture (in the sense that some
of the above mentioned aspects could be missing), the results reported below
suggest that it captures striking phenomena such as the impact of screening
length on capillary freezing. Among its strengths, this method allows
considering nearly any screening length but also any electrode shape; indeed,
by using a liquid phase of screening charges within the solid material, solid
surfaces with complex morphology can be considered (this contrasts with other
approaches which require defining an underlying atomic structure bearing
localized polarizable electrons). Moreover, as it can be directly implemented
in any standard molecular dynamics package, our approach can be used for
fluids ranging from simple charged/dipolar molecules to more complex systems
such as ionic liquids.

	\begin{figure}[htbp]
    \begin{center}
        \includegraphics[width=.8\columnwidth]{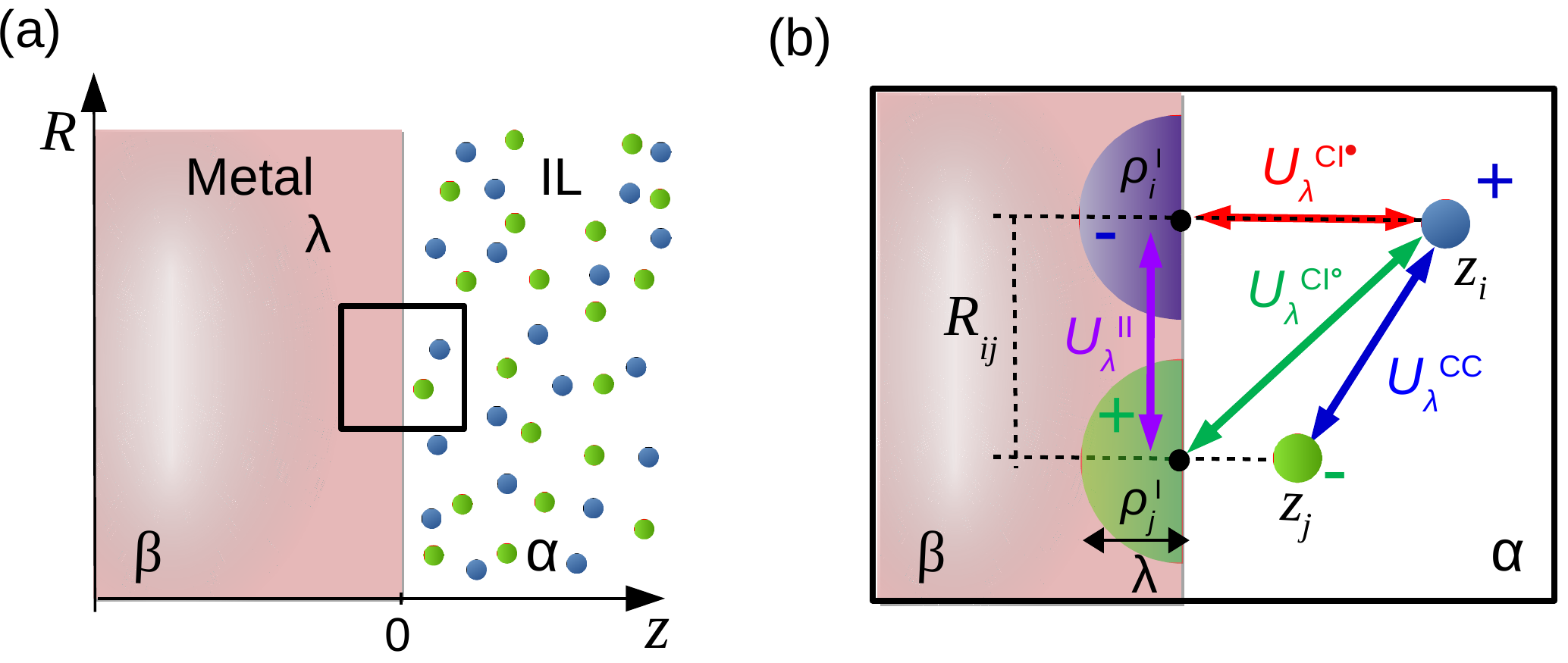}
    \end{center}
    \caption{\textbf{Electrostatic interactions in the vicinity of metal
            surfaces.} (a) Ionic liquid (IL) in an insulating medium $\alpha$
            close
        to an
        imperfect metal $\beta$ having a Thomas--Fermi length
        $\lambda$. (b) One and two-body interactions for two point charges
        $i,j$ at distances $z_i$ and $z_j$ from the surface and separated
        by in-plane distance $R_\mathit{ij}$. The induced charge
        distribution $\rho_k^\mathrm{I}(\mathbf{r})$ for $k = i,j$ (denoted
        by  half-ellipsoids) within the metal is of opposite sign and
        decays
        over $\lambda$. The colored arrows show the different energy
        contributions given in \cref{eq:ion_energy_ij}.}
    \label{fig:fig_1}
\end{figure}

\noindent	\textbf{Interaction at Thomas--Fermi metal interfaces}\\
\noindent	Fig.\ 1(a) depicts point charges in an insulating
medium $\alpha$
of relative dielectric constant $\varepsilon_\alpha$ close to a metal
$\beta$ of TF length $\lambda$.
As shown in \textit{Supplementary Information II D}, the electrostatic
energy of two charges $i$ and $j$ at distances $z_i$ and $z_j$ from the
dielectric/metal interface and separated by $r_{ij} = \lbrack R_{ij}^2+(z_i
- z_j)^2 \rbrack^{1/2}$ with $R_{ij}$ the in-plane distance reads:
\begin{equation}
    U_{\lambda}(z_i,z_j,R_{ij}) =
    U^\mathrm{CC}(r_{ij}) + U_{\lambda}^\mathrm{CI}
    (z_i,z_j,R_{ij})
    + U^\mathrm{II}_{\lambda}(z_i,z_j,R_{ij}).
    \label{eq:ion_energy_ij}
\end{equation}
where the superscripts C and I refer to the physical charges in the
dielectric medium and induced charges within the metal, respectively. As
shown in Fig.\ 1(b), $U^\mathrm{CC}$ is the Coulomb interaction
energy between the charges $i$ and $j$ while $U^\mathrm{II}_{\lambda}$ is
the interaction energy between the charge densities $\rho_i^\mathrm{I}$ and
$\rho_j^\mathrm{I}$ induced in the metal by these two charges. For each ion
$i$, its interaction energy $U_{\lambda}^\mathrm{CI}$ with the metal
decomposes into a one-body
contribution $U_{\lambda}^{\mathrm{CI}^\bullet}(z_i)$ --- corresponding to
the interaction with its image in the metal --- and two-body contributions
$U_{\lambda}^\mathrm{CI^\circ} (z_i,z_j,R_{ij})$ --- corresponding to the
interaction with the induced charges due to all other charges $j$.
Analytical expressions exist for $U_{\lambda}^{\mathrm{CI}^\bullet}$ and
$U_{\lambda}^\mathrm{CI^\circ}$\cite{newns_1969_fermi,
    inkson_1973_manybody,kornyshev_1977_image,kaiser_2017_electrostatic}
but	 $U^\mathrm{II}_{\lambda}$ must be estimated
numerically from the energy density, i.e. $U^\mathrm{II}_{\lambda} =
\int\rmd\mathbf{r} \Psi^\mathrm{\beta}_{i}(\mathbf{r})
\rho^\mathrm{I}_{j}(\mathbf{r})	+
\Psi^\mathrm{\beta}_{j}(\mathbf{r})\rho^\mathrm{I}_{i}(\mathbf{r})$,
where $\Psi^\mathrm{\beta}_{k}$ and $\rho^\mathrm{I}_{k}$ are the
electrostatic potential and induced charge density in the metal due to the
point charge $k = i, j$. All details are given in the \textit{Supplementary
    Information}.

	\noindent	\textbf{Effective molecular simulation approach} \\
\noindent Except for the usual Coulomb energy CC, formal expressions for
the CI and
II energies cannot be implemented in molecular simulation due to their
complexity. In particular, $U^\mathrm{II}_{\lambda}$ requires expensive
integration on the fly as analytical treatment for imperfect metals is only
available in closed forms in asymptotic
limits.\cite{netz_1999_debyehuckel,kaiser_2017_electrostatic}
Here we model the resulting complex electrostatic interactions between the
ions of the liquid thanks to a `virtual Thomas--Fermi fluid' located within
the confining solids, see Fig.\ 2(a).
Our approach builds on the direct analogy between the Thomas--Fermi
screening of electrons and the Debye--H\"uckel equation for electrolyte
solutions.
In the linear Thomas--Fermi formalism, the induced electronic charge
density in the metal writes:
$q_\mathrm{TF} \rho^\mathrm{I}(\mathbf{r}) = -\varepsilon_0
\varepsilon_\beta k_\mathrm{TF}^2 \Psi_\beta(\mathbf{r})$
where $\varepsilon_0$ is the vacuum permittivity,
$\varepsilon_\beta$ the relative dielectric constant,
$k_\mathrm{TF} =\lbrack
e^2\mathcal{D}(\mathcal{E}_\mathrm{F})/\varepsilon_0
\varepsilon_\beta \rbrack^{1/2}$ the Thomas--Fermi wave-vector, and
$\mathcal{D}(\mathcal{E}_\mathrm{F})$ the density of states at the Fermi
level (see \textit{Supplementary Information II A}).
Combined with Poisson equation, this leads to the Helmholtz
equation for TF screening, $\nabla^2 \Psi_\mathrm{II} =
k_\mathrm{TF}^2\Psi_\mathrm{II}$,
which indeed resembles the Debye--H\"uckel equation for electrolyte
solutions.
Accordingly, one can simulate the imperfect metal using a system of
virtual (classical) charged particles of
charge $q_\mathrm{TF}$ and mass $m_\mathrm{TF}$, with
density $\rho_\mathrm{TF}$ and temperature $T_\mathrm{TF}$.
The analogous TF screening length $\lambda = k_\mathrm{TF}^{-1}$
can be identified as the equivalent Debye length $\lambda_\mathrm{D}$:
\begin{equation}
    \lambda \sim \lambda_\mathrm{D} =
    \sqrt{\frac{\varepsilon_\beta \varepsilon_0
            \kbT_\mathrm{TF}}{\rho_\mathrm{TF} q_\mathrm{TF}^2}}.
    \label{eq:ktf}
\end{equation}
Hence, by considering the dynamics 
of these light ions located in the confining solid, any screening length
$\lambda$ between 0 (perfect metal) and $\infty$ (insulator) can be
efficiently mimicked depending on $q_\mathrm{TF}$, $\rho_\mathrm{TF}$, and
$T_\mathrm{TF}$.
This virtual system allows simulating the complex electrostatic
interactions within the ionic liquid in the vicinity of an imperfect metal.

As mentioned in the previous paragraph,
by tuning the different parameters inherent to the Thomas--Fermi fluid ---
namely,
the fluid particle charge $q_\mathrm{TF}$, temperature $T_\mathrm{TF}$ and
density $\rho_\mathrm{TF}$ --- confining solids with an electrostatic
screening
length ranging from metallic to insulating can be mimicked.
Yet,  \cref{eq:ktf} shows that mapping the fluid of
mobile charges onto the TF model only requires to set a single parameter
$\rho_\mathrm{TF} q_\mathrm{TF}^2/T_\mathrm{TF}$ (fixing
$\varepsilon_\beta=1$).
In practice, while this implies that  different combinations for these
parameters
can lead to the same effective screening, there are a few constraints that
should be verified.
First, as discussed hereafter in this paragraph, to account for the ultra-fast
dynamical relaxation of charges in the solid compared to that in the confined
salt,
the Thomas--Fermi temperature $T_\mathrm{TF}$ is chosen to be very large.
While this is important to capture the order of magnitude difference between
these two timescales,
this implies that two thermostats have to be used to properly regulate the
temperature
of the two subsystems. In this regard, we emphasize that the results reported
here were obtained using either a Nose-Hoover thermostat or a Langevin
thermostat (however, it was found that a Langevin thermostat is recommended
as it ensures that the two subsystems display uncoupled homogeneous/constant
temperatures). Moreover, in contrast to the temperature of the confined
charges,
$T_\mathrm{TF}$  should not be seen as a physical temperature but rather as a
parameter
governing the Thomas--Fermi fluid dynamics and, hence, effective screening.
Second, since the ionic force in the Debye length is directly the product
of the squared charge and density, i.e. $\rho_\mathrm{TF} q_\mathrm{TF}^2$,
one can tune the effective screening in the metal by tuning either one or two
of these parameters.
In practice, after conducting several tests, we found it more efficient
to keep the number and, hence, the density of charges in the Thomas--Fermi
fluid constant.
Indeed, on the one hand, playing with $q_\mathrm{TF}$ at constant
$\rho_\mathrm{TF}$ offers
more flexibility in tuning the screening length. On the other hand,
treating very imperfect metals with constant $q_\mathrm{TF}$ would require
considering a very small number of Thomas--Fermi ions leading to poor
sampling/statistics performance.
Before going into more technical details, we emphasize that the exact
choice of parameters made to mimic different screening lengths is expected
to impact the dynamics/kinetics of the observed phenomena
(by setting a given temperature $T_\mathrm{TF}$, we do impose a relaxation
time).
However, like in classical thermodynamics, we do not expect this effective
relaxation within the virtual Thomas--Fermi fluid to affect the equilibrium
properties
of the charged liquid confined between the metallic surfaces.

In more detail, in our molecular dynamics approach,
to ensure that the particles in the effective Thomas--Fermi fluid relax fast,
their mass/temperature are chosen much smaller/larger than their counterpart
in the confined system; typically $m_\mathrm{TF} \sim 0.01 m$ and
$T_\mathrm{TF} \sim 10 T$ (requiring typical integration steps of 0.1 fs and
1 fs, respectively).
In practice, as shown in Fig.\ 2(a), the
effective simulation strategy consists of sandwiching the charged or
dipolar system between two metallic media separated by a distance
$d_\mathrm{w}$. The confining media of width $d_\mathrm{TF}$
are filled with the Thomas--Fermi fluid having a density
$\rho_\mathrm{TF}$. Once
$\rho_\mathrm{TF}$ and $T_\mathrm{TF}$ are set, $\lambda_\mathrm{D}$ is
varied by
tuning $q_\mathrm{TF}$ according to \cref{eq:ktf}; from $q_\mathrm{TF} = 0$
$(\lambda_\mathrm{D} \rightarrow \infty$) for an insulator to $q_\mathrm{TF}
= 1$
($\lambda_\mathrm{D} =0.03\,\mathrm{nm}$) for a nearly perfect metal.
All simulations reported in this article are carried out using molecular
dynamics but they could be easily implemented into Monte Carlo algorithms
to perform calculations in other ensembles (Grand Canonical and
isothermal/isobaric ensembles for instance).
All simulation details are provided in the Methods section.
Since we consider a set of explicit charges with no solvent to model the
Thomas--Fermi fluid, we used consistently a relative dielectric permittivity
equal to 1.
However, this raises the question of the true dielectric constant of the
solid material that we intend to mimic.
Describing \textit{s}-\textit{p} metals using a Thomas--Fermi model is
usually done by accounting for the dielectric constant of the underlying ion
structure in the metal.
Such dielectric background, which arises from the polarizability of the core
electron shells around each metal atom, differs from one metal to another
(with values of the order of one to a few times the vacuum dielectric
permittivity).
In semiconductors, such a dielectric background, in which the outer electrons
move, arises from interband transitions with values that can be close to 10.
As another example, when modeling semi-metals, a large Thomas--Fermi
screening length is used together with a large dielectric constant to account
for the low charge carrier concentration and absence of band-gaps,
respectively.
In this context, we mention that the dielectric constant in semi-metals can
be evaluated by considering the electron density of states as proposed by
Gerischer and
coworkers.\cite{gerischer_1985_interpretation,gerischer_1987_density}
These specific yet representative examples illustrate that the nature of the
solid material manifests itself in the Thomas--Fermi screening length but
also in the dielectric constant. While this duality (screening/dielectric
properties) is of course of prime importance when modeling the complex
physics of metallic and semiconducting surfaces, we simplify the problem in
the present approach as we only aim at mimicking in an effective but
quantitative fashion the electrostatic screening induced by the solid surface
on the confined or vicinal charges.
To this end, as already introduced above, our proposed molecular approach
adopts a different view by mimicking a liquid of charges to produce an
effective screening that corresponds to an underlying Thomas--Fermi model ---
this is what is referred to in our article as `virtual' Thomas--Fermi fluid.
In other words, as will be shown below, by calculating the induced
electrostatic screening length, we can map this pseudo Thomas--Fermi medium
onto practical situations by establishing a consistent relation between its
Debye length $\lambda_\textrm{D}$ and the induced electrostatic screening
length $\lambda$. Therefore, we are not attempting \textit{per se} to
implement the electrostatic screening equations as derived using the
Thomas--Fermi formalism. In this regard, as will be established later, we
believe that our mapping procedure is sound and robust as the inferred
screening length is found to be consistent with that corresponding to the
observed capacitance behavior of our system.

	\begin{figure}[htbp]
		\begin{center}
			\includegraphics[width=.8\columnwidth]{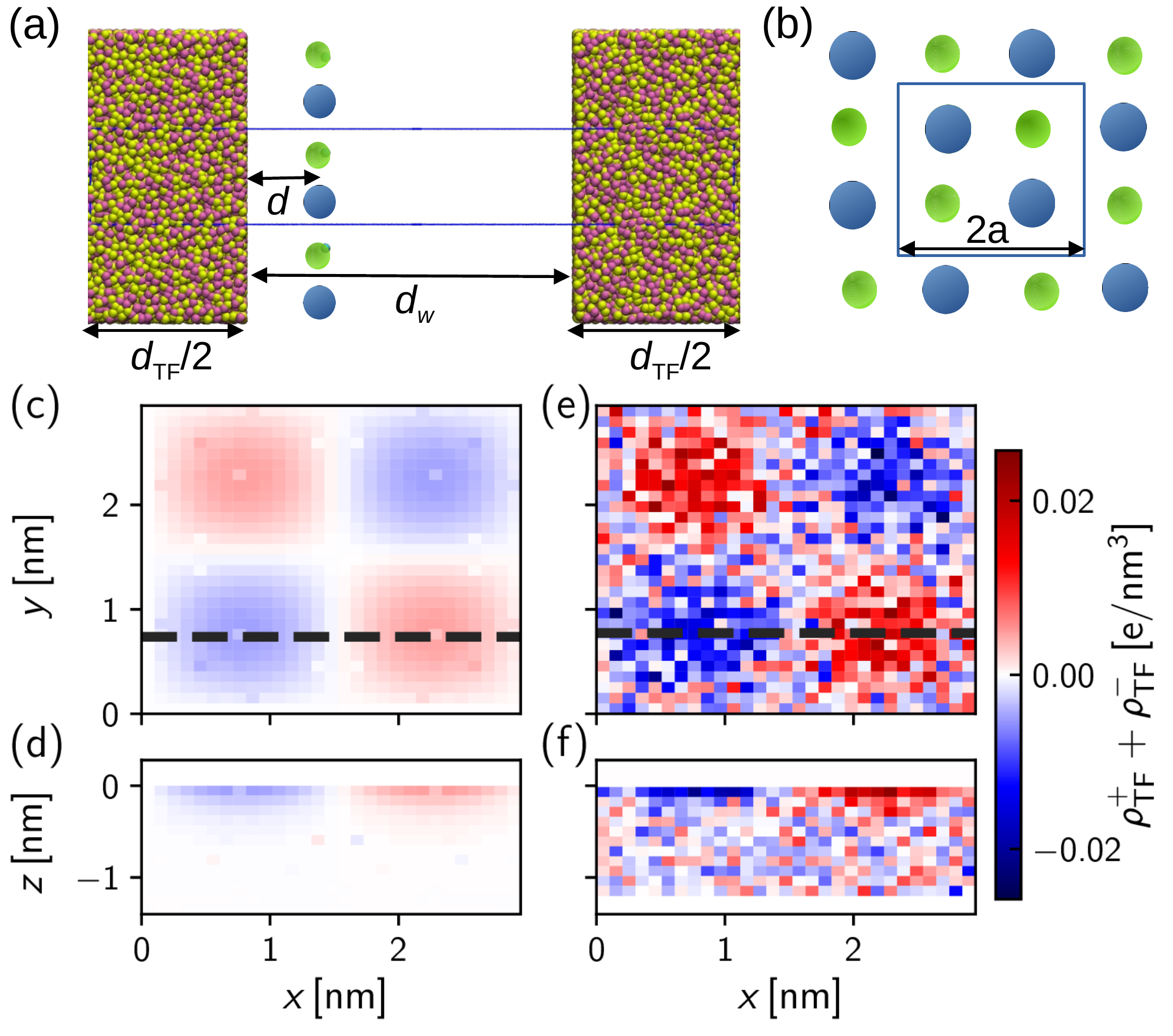}
		\end{center}
		\caption{\textbf{Induced charges and screening at metallic surfaces.}
			(a) 2D ionic crystal (blue/green charges) at a distance
			$d=0.22\,\mathrm{nm}$ from a medium in which electrostatic
			screening is modeled using a Thomas--Fermi fluid (yellow/pink
			charges). The crystal layer is confined in a pore of width
			$d_\mathrm{w}$ 	while the Thomas--Fermi fluid occupies a region
			of width $d_\mathrm{TF}$. (b) Top view of the 2D ionic crystal
			(lattice
			constant $a$) illustrating the periodic boundary conditions.
			(c)-(d) Top/side views of the induced charge density
			$\rho^\mathrm{I}(d,\mathbf{r})$ in a Thomas--Fermi fluid as
			obtained
			from \cref{eq:rhoind} for the system in (a).
			(e)-(f) Same as (c)-(d) but using our simulation approach.
			The screening length as defined in \cref{eq:ktf} is
			$\lambda=0.25\,\mathrm{nm}$.
		}
		\label{fig:fig_2}
	\end{figure}

	\noindent	\textbf{2D crystal at metallic interfaces} \\
\noindent	To validate our effective approach, we consider a 2D square
crystal of
lattice constant $a=1.475\,\mathrm{nm}$ made up of charges $\pm
1\,\mathrm{e}$ and located at a distance $d$ from a metal
(Fig.\ 2). Due to the periodic boundary conditions, a second
pore/metal interface is present at a distance
$d_\mathrm{w}=20\,\mathrm{nm}$. Yet, as shown in \textit{Supplementary
    Figure S9}, this second interface does not affect the electrostatic
energy
as $d_\mathrm{w}$ is large enough. In the Thomas--Fermi framework, the
charge density $\rho^\mathrm{I}$ at a position $\mathbf{r}$ in the metal
induced by a charge $q$ located in $(0,0,d)$ reads (see
\textit{Supplementary Information II D}):
\begin{equation}
    \rho^\mathrm{I}(d,z,R) = -\int\limits_0^\infty \rmd K K J_0(KR)
    \frac{\varepsilon_\beta k_\mathrm{TF}^2 q \rme^{-K d}}
    {2\pi\left(\varepsilon_\alpha K +
        \varepsilon_\beta \kappa \right)}
    \rme^{\kappa z},
    \label{eq:rhoind}
\end{equation}
where $R=\lbrack x^2 + y^2 \rbrack^{1/2}$ is the lateral distance to the
charge $q$, $J_0$ is Bessel function of the first kind, and
$\kappa^2=K^2+k_\mathrm{TF}^2$. Figure 2(c,d) shows the induced
charge density $\rho^\mathrm{I}(d,\mathbf{r})$ as obtained by summing
\cref{eq:rhoind} for the 2D crystal when $d=0.22\,\mathrm{nm}$ and
$\lambda_\mathrm{D} = k_\mathrm{TF}^{-1}=0.25\,\mathrm{nm}$
(as discussed in \textit{Supplementary Information III},
\cref{eq:rhoind} must be summed over all
crystal periodic images but it was found that the sum converges quickly).
For comparison, Fig.\ 2(e,f) shows
$\rho^\mathrm{I}(d,\mathbf{r})$ as obtained using our effective approach
from the local charge density in the metal, i.e.\ $\rho^\mathrm{I} =
e(\rho_\mathrm{TF}^+ - \rho_\mathrm{TF}^-)$.
In contrast to $\rho^\mathrm{I}(d,\mathbf{r})$ in the Thomas--Fermi model,
due to their finite size, the fluid charges in the simulation cannot
approach arbitrarily close to the metal/pore surface. For consistency, the
analytical/simulation data were compared by defining $z = 0$ in the
simulation as the position where the Thomas--Fermi fluid density becomes
non-zero. Fig.\ 2 shows that the effective molecular simulation
qualitatively captures the predicted density distribution induced in the
metal. Each physical charge in the 2D crystal induces in the
metal a diffuse charge distribution of opposite sign. Moreover, as expected
from the Thomas--Fermi framework, the induced charge distribution in the
effective simulation decays over the typical length $\lambda_\mathrm{D}$.

Our effective approach was assessed quantitatively by probing the energy of
the 2D ionic crystal as a function of its distance $d$ to the metal surface
for different screening lengths $\lambda$. The simulated electrostatic
energy  $U_{\lambda}(d)$ consists of all ion pair contributions in
\cref{eq:ion_energy_ij} as discussed in \textit{Supplementary
    Information V}.
Figure 3 compares the total energy $U_\lambda$ as a
function of the distance $d$ with the numerically evaluated prediction from
\cref{eq:ion_energy_ij}. As expected
theoretically, the overall energy decays with decreasing $\lambda$
between boundaries for an insulator ($\lambda \to \infty$) and a perfect
metal ($\lambda \to 0$).
As shown in Fig.\ 3, our effective approach captures
quantitatively the screening behavior of the confining medium assuming a
screening length
$\lambda = c_0 + c_1\lambda_\mathrm{D} + c_2\lambda_\mathrm{D}^2$
(with $\lambda_\mathrm{D}$ the ion gas 	Debye length,
$c_0=0.22\,\mathrm{nm}$, $c_1=0.91$ and $c_2=0.28\,\mathrm{nm}^{-1}$ in our
system).
Such values do not simply correspond to fitting parameters that allow
matching the
simulated and theoretical energies; as explained in the next paragraph,
they were derived so that the capacitance of the virtual Thomas--Fermi
fluid matches the theoretically expected value $C = \varepsilon_0/\lambda$.
The fact that the rescaled screening length $\lambda$ also allows
recovering the expected screened interaction
energy further supports the physical validity of our effective molecular
approach. Moreover, physically, the parameters $c_0$, $c_1$, and $c_2$ are
not just empirical parameters as they account for the following effects in
the screening fluid used in the simulation:
$c_0$ accounts for the finite size $\sigma$ of the Thomas--Fermi
ions which prevents reaching screening $\lambda \leq \sigma$. This is
supported by the fact that $c_0 \sim \sigma$ corresponds to the value below
which the repulsive interaction potential in the Thomas--Fermi fluid becomes
larger than $k_\textrm{B}T$.
$c_1$ arises from the non-ideal behavior of the effective
Thomas--Fermi fluid which leads to overcreening compared to an ideal gas
having
the same charge density $\rho_\mathrm{TF}$ ($c_1 = 1$ corresponds to the
ideal behavior);
$c_2 \neq 0$ indicates non-linear effects in electrostatic screening which
go beyond the linear approximation used in the Thomas--Fermi framework.

To illustrate the ability of our approach to capture the impact of various
screening lengths --- from insulators to metallic surfaces --- on
electrostatic interactions, we show in \textit{Supplementary Fig.~S1(a)} the
electrostatic energy arising from image/charge interactions,
$U^\mathrm{CI}(\lambda)$, for a molten salt confined between two solid
surfaces as a function of the electrostatic screening length $\lambda$. To
probe the impact of such interactions with induced charges, a fixed liquid
configuration was considered as it implies that the direct coulomb
interaction  $U^\mathrm{CC}(\lambda)$ is constant. As expected, upon
decreasing $\lambda$, the overall electrostatic energy decreases as the
interactions with the induced charges in the metallic surfaces become more
negative. Moreover, by extrapolating $U^\mathrm{CI}(\lambda)$ to perfect
metallic conditions (i.e. $\lambda \to 0$), one recovers the expected charge
image contribution corresponding to half of the Coulomb energy. Such data
show that our molecular simulation strategy does mimic --- albeit in an
effective fashion --- the electrostatic screening induced by metallic
surfaces. In this respect, we emphasize that this approach can be extended to
almost any surface geometry/topology. First, this versatile model does not
require inputting an underlying atomic structure for the surface (by
describing the charges in the metal as a fluid, one needs not to consider an
atomic lattice to which charges are linked). Second, any geometry from a
simple flat or cylindrical surface to disordered/rough surfaces can be
considered as it simply requires to encapsulate the screening charges within
a mathematically defined region.
Correctly accounting for image forces near solid surfaces is crucial to
capture the rich and complex behavior of confined charges. In particular, as
theoretically predicted by Kondrat and
Kornyshev,\cite{kondrat_2010_superionic} it has been observed using molecular
simulation that such image forces can lead to superionic states --- where
like-charge pairs form --- in metallic
nanoconfinement.\cite{li_2018_computer} In this respect, it was found that
electron spillover leads to effective pore sizes narrower than the nominal
pore size. While the Thomas--Fermi model was found to fail to predict this
effective pore size
reduction,\cite{rochester_2013_interionic,mohammadzadeh_2016_energetics} we
believe that --- at least as a first order approach --- the results reported
in the present paper show that our molecular strategy can be used to capture
the impact of electrostatic screening on ions between metallic surfaces.
In particular, by comparing molecular simulation for an ionic liquid on a
perfect metal surface at constant potential and constant charge, it was
recently shown that image forces are screened on such a very short scale that
they do not affect the adsorbed liquid.\cite{bi_2018_minimizing}
This highlights the need to consider effective molecular simulation
approaches such as the one reported in this paper to consider imperfect
metals and, hence, larger screening lengths corresponding to many
experimental situations.

	\begin{figure}[htbp]
		\begin{center}
			\includegraphics[width=.6\columnwidth]{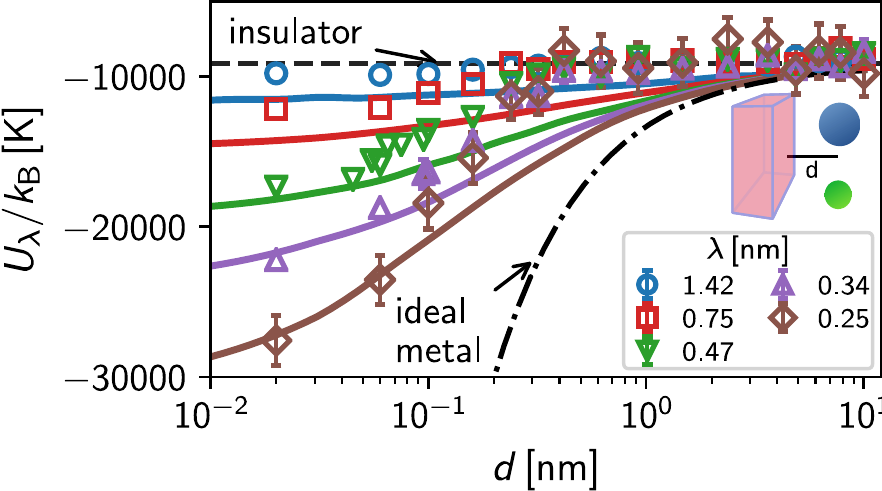}
		\end{center}
		\caption{\textbf{Screened electrostatic interactions through the use of
				a virtual Thomas--Fermi fluid.}
			Electrostatic energy  $U_\lambda(d)$ between a 2D ionic
			crystal and a Thomas--Fermi metal separated by a distance $d$ for
			different $\lambda$. For each $\lambda$, the symbols correspond to
			the effective simulation while the solid line shows the linear
			Thomas--Fermi predictions.}
		\label{fig:fig_3}
	\end{figure}

	\noindent \textbf{Capacitance}\\
\noindent To further establish the validity of our novel molecular
approach, an important requirement is to verify that the virtual
Thomas--Fermi fluid yields the correct capacitance behavior. With this aim,
as shown in Fig.\ 4(a), a simple molecular dynamics set-up was
designed by assembling a composite system made up of two Thomas--Fermi
fluids sandwiching a dielectric material of a width $d_\mathrm{w}$ (either
a vacuum layer or a molten salt was considered to verify that the overall
capacitance follows the expected physical behavior).
The molten salt (NaCl) is modeled using charged particles
$\pm 1\rme$ that interact via a Born–-Mayer–-Huggins
potential\cite{anwar_2002_calculation} (see
\textit{Supplementary Table S1}).
To prevent mixing of the
Thomas--Fermi fluid/charged system, a reflective wall of thickness $\xi =
0.2$ nm is positioned between the two subsystems.
The whole composite is placed between two electrodes having an overall
charge $+Q$ and $-Q$ (all details can be found in the Methods section).
With such a geometry, the capacitance $C=Q/\Delta \Psi$ is readily obtained
from the potential difference $\Delta \Psi$. As shown in
Fig.\ 4(a), with this molecular simulation set-up mimicking in a simplified
yet realistic way an experimental electrochemical cell, we can perform a
molecular dynamics simulation to readily estimate the positive and negative
charge density profiles within the confined salt [$\rho_+(z)$ and
$\rho_-(z)$]. Using Poisson equation, i.e.\ $\Delta \Psi(z) =
-\rho(z)/\epsilon$ with  $\rho(z) = e[\rho_+(z) - \rho_-(z)]$, $\Delta
\Psi(z)$ is determined by integrating twice the charge density profile
$\rho(z)$. Fig.\ 4(a) shows $\Delta \Psi(z)$ as a function of the position
$z$ within the confined liquid for different screening lengths $\lambda$. In
practice, two simple situations are considered; the porosity between the two
metallic surfaces is either occupied by vacuum or by a molten salt. As
expected, $\Delta \Psi$ increases with increasing $z$ as the negative
electrode is located at $z = 0$ (positive charge adsorption) and the positive
electrode is located at $z = d_\mathrm{w}$ (negative charge adsorption).
Moreover, by considering the data sets for vacuum-filled and liquid-filled
pores in  Fig.\ 4(a), we observe that the slope of $\Delta \Psi(z)$ in the
pore region is larger for the former than for the latter. Considering that
$C=Q/\Delta \Psi$, this result suggests that as expected the sandwiched salt
layer has a larger capacitance than the sandwiched vacuum layer. To provide a
more quantitative picture of the system capacitance as a function of the
screening length $\lambda$, we performed in the following paragraph a more
detailed analysis in which the capacitance of the different elements ---
confined material and Thomas--Fermi fluid --- is extracted.

The system considered here simply consists of double layer  capacitors in
series so that its capacitance per
unit area should verify the following combination rule:
\begin{equation}
    \frac{1}{C} = \frac{1}{C_\mathrm{vac}} + \frac{2}{C_\mathrm{TF}} =
    \frac{d_\mathrm{w}+2e}{\varepsilon_0} + \frac{2\lambda}{\varepsilon_0},
\end{equation}
where the first and second terms correspond to the capacitance of the
vacuum slab of
width $d_\mathrm{w}$ and that of the Thomas--Fermi fluid (the factor 2
simply accounts for the presence of two Thomas--Fermi/vacuum interfaces). As
shown in Fig.\ 4(b),  the simulation data are in reasonable
agreement with the prediction from this simple expression with deviations
increasing with $\lambda$. Interestingly, as shown in the insert in
Fig.\ 4(b), our effective approach captures quantitatively the
expected capacitance behavior of the confining medium upon  rescaling
$\lambda_\mathrm{D}\to\lambda = c_0 + c_1\lambda_\mathrm{D} +
c_2\lambda_\mathrm{D}^2$
(see discussion above).
As another consistency check, the vacuum layer in the capacitor was
replaced by a slab of molten salt --- see molecular configuration shown in
Fig.\ 4(a).
As expected, upon inserting such a molten salt, the effective capacitance
$C$ drastically increases (i.e.\ the inverse capacitance shown in
Fig.\ 4(b) decreases). More importantly, as shown in
Fig.\ 4(c), the induced capacitance change $\Delta 1/C$ observed
in our simulation data follow the expected behavior with a
$\lambda$-independent value:
\begin{equation}
    \Delta 1/C =
    d_\mathrm{w}\frac{\varepsilon-\varepsilon_0}{\varepsilon\varepsilon_0},
\end{equation}
where $\varepsilon$ is the permittivity of the molten salt.
Furthermore, since $\varepsilon\gg\varepsilon_0$, we predict that $\Delta
1/C \sim d_\mathrm{w}/\varepsilon_0$ in very good agreement with the
simulation data shown as green circles in Fig.\ 4(c) (the small
deviation is due to the fact that the vacuum permittivity is not completely
negligible compared to that of the molten salt).

	\begin{figure}[htbp]
		\begin{center}
			\includegraphics[width=\textwidth]{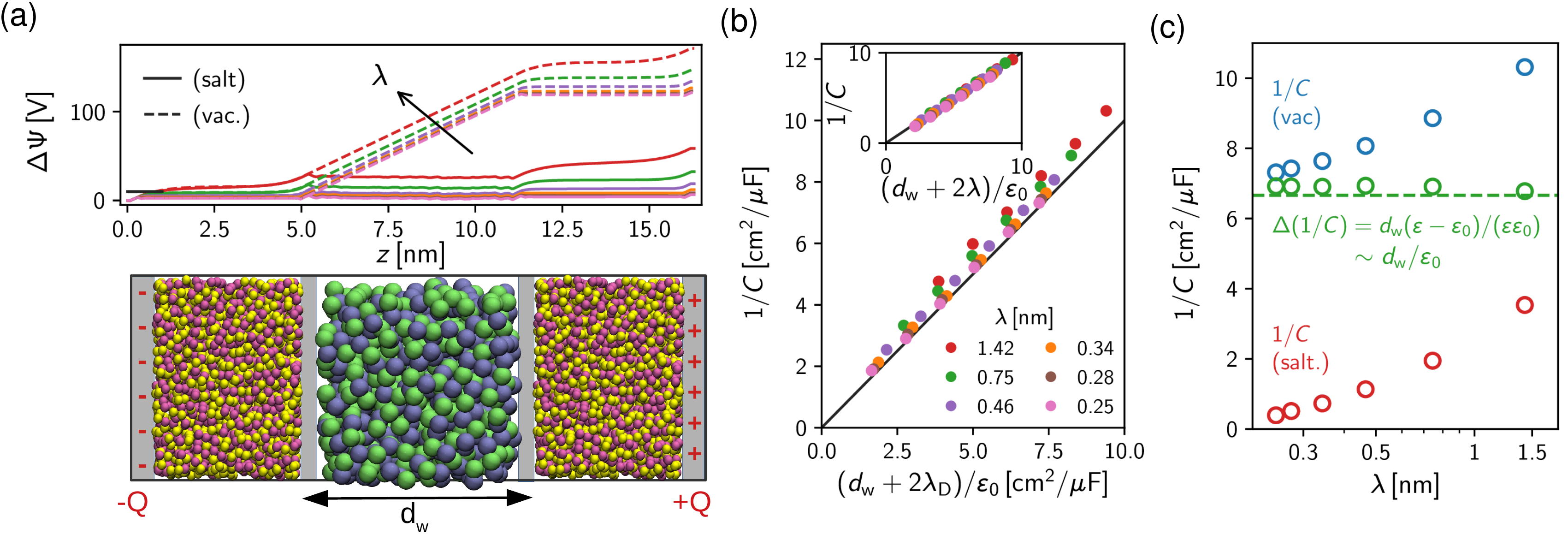}
		\end{center}
		\caption{
    \setlength\parfillskip{\leftskip}
    \textbf{Capacitive behavior of virtual Thomas--Fermi fluids.} (a)
    Typical molecular configuration of the molecular dynamics set-up
    employed to determine the capacitive behavior of the virtual
    Thomas--Fermi fluid. A molecular system --- here, a molten salt
    corresponding to the blue/green spheres --- is confined between two
    virtual Thomas--Fermi fluids (yellow/pink spheres). This composite
    system is sandwiched by two electrodes having a constant surface
    charge $\pm Q$. As shown in the top figure, using this set-up,  the
    electrostatic potential
    profile $\Delta \Psi(z)$ can be determined by integrating twice the
    resulting charge  density profile $\rho(z) = e[\rho_+(z) -
    \rho_-(z)]$. Such numerical assessments were performed for
    different screening lengths $\lambda$ where the confined material
    is either a vacuum layer (dashed lines) or  a molten salt (solid
    lines).
    (b) Reciprocal capacitance $1/C$ of the empty Thomas--Fermi
    capacitor for different $\lambda_\mathrm{D}$ versus the analytical
    prediction for two double layer capacitors in series.
    The inset shows that the simulation data collapse onto the same
    master curve when plotted using the effective screening length
    $\lambda$.
    (c) Reciprocal capacitance $1/C$ as a function of the screening
    length $\lambda$ for a capacitor made up of vacuum (red symbols) or
    molten salt
    (blue symbols) confined between the TF fluids. As expected, for
    $\varepsilon=\varepsilon_\alpha\varepsilon_0 \gg \varepsilon_0$,
    the difference between both systems (green symbols) is close to
    $d_\mathrm{w}/\varepsilon_0$.
}
\end{figure}

\noindent \textbf{Capillary freezing/melting in confinement}\\
\noindent
In what precedes, our effective molecular approach was shown to capture the
electrostatic energy predicted using the Thomas--Fermi formalism for
electrostatic screening in metallic materials as well as the capacitive
behavior of a molten salt sandwiched between metallic surfaces with different
screening lengths. Yet, in addition to these two validation steps, it should
be verified that our simple strategy allows  reproducing available
experimental data for realistic materials. To do so, we have performed
additional calculations using our effective treatment to study the
liquid/crystal phase transition in various metallic confinements as
experimentally reported by Comtet et al.\cite{comtet_2017_nanoscale} By
considering the crystallization of an ionic liquid confined between an AFM
tip and a metallic surface, these authors showed that the melting temperature
is shifted above the bulk melting point and that the shift in the melting
point increases with decreasing the screening length. To help rationalize
these results, we performed the following molecular simulation study using
our effective electrostatic screening strategy in confined charged systems.

\begin{figure}[htbp]
    \begin{center}
        \includegraphics[width=.9\columnwidth]{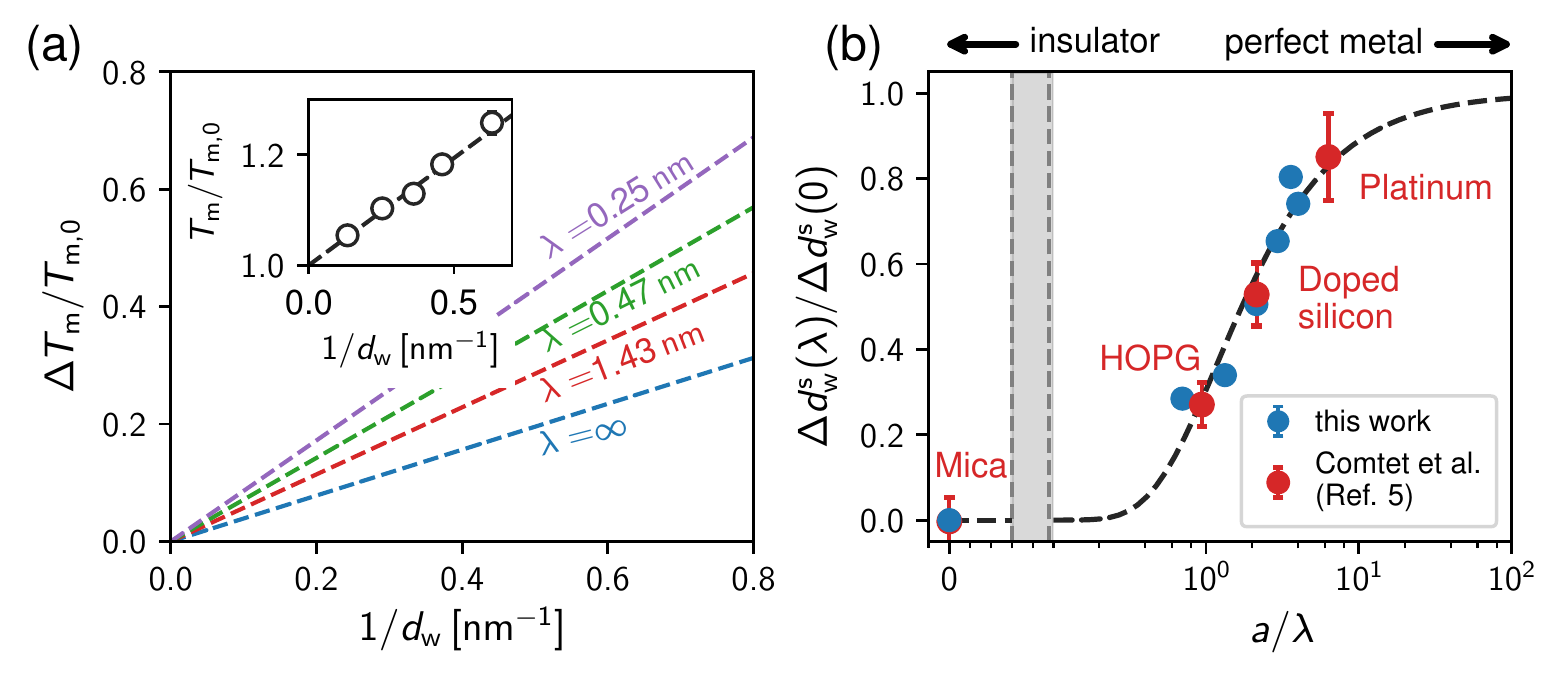}
    \end{center}
    \caption{\textbf{Capillary freezing at metallic surfaces.}
        (a) Shift in the melting point $\Delta T_m$ with respect to the bulk
        melting temperature $T_{\mathrm{m},0}$ for a salt confined between
        two flat surfaces separated by a distance $d_\textrm{w}$. The
        confining surfaces are made up of different types of materials
        characterized by their electrostatic screening length $\lambda$: from
        insulator  ($\lambda \to \infty$) to more and more perfect metals
        ($\lambda =$ 1.43, 0.47, and 0.25 nm). The insert shows the melting
        point $T_\mathrm{m}$ normalized to the bulk melting point
        $T_{\mathrm{m},0}$ as a function of the reciprocal pore size
        $d_\textrm{w}^{-1}$ as obtained from the direct coexistence method
        for insulating surfaces. (b) Shift induced in the capillary freezing
        length
        $\Delta
        d_\textrm{w}^\textrm{s}(\lambda)=d_\textrm{w}^\textrm{s}(\lambda)-d_\textrm{w}^\textrm{s}(\infty)$
        for a salt confined between two surfaces characterized by their
        electrostatic screening length $\lambda$. The data are plotted with
        respect to  the value obtained for an insulating surface
        $d_\textrm{w}^\textrm{s}(\infty)$. The blue circles are simulation
        data for a simple molten salt while the red circles	are the
        experimental data by Comtet et al.\ who considered the capillary
        freezing of a room temperature ionic liquid [BMIM]/[BF4] confined
        between different materials as labeled in the figure.
        The dashed line corresponds to the predictions obtained by fitting
        the data $\Delta \gamma(\lambda)$.}
    \label{fig:fig_Tm}
\end{figure}

First, by considering insulating surfaces, we use the direct coexistence
method (DCM) in which a crystalline salt coexists with its molten salt in a
slit pore of a size $d_\mathrm{w}$ (various pore widths between 1.5\ and 7.1
nm were considered).
To mimic a physical system in which the confined phases also interact through
simple dispersive/repulsive interactions with the surface, a 9-3
Lennard-Jones interaction potential was added between each salt atom and the
solid surface.
The use of such structureless surfaces to describe the confining solids was
made to avoid inducing a peculiar crystalline structure by employing a given
molecular periodic lattice.
Moreover, to ensure that such dispersive interactions do not impact
too much the melting point in confinement, the potential wall depth
was chosen of the order of $k_\textrm{B}T$ and, hence, at a value
much lower than the electrostatic interactions.
Using this set-up, molecular dynamics simulations in the canonical ensemble
are performed for different $T$ to determine the melting temperature
$T_\mathrm{m}$ as follows.
The crystalline salt melts into the liquid phase for $T>T_\mathrm{m}$  while
the molten salt crystallizes into the crystal phase for $T < T_\mathrm{m}$.
The insert in Fig.\ 5(a) shows the shift in the melting point
with respect to the bulk melting point, $\Delta T_\mathrm{m} = T_\mathrm{m} -
T_\mathrm{m,0}$
as a function of pore size $d_\mathrm{w}$.
In agreement with the experimental data,\cite{comtet_2017_nanoscale} these
data show that the salt confined between insulating surfaces has a melting
temperature above the bulk melting point.
Moreover, the melting point shift $\Delta T_\mathrm{m}/T_\mathrm{m,0}$ is
found to scale with the reciprocal pore size $1/d_\mathrm{w}$ as predicted
using the Gibbs-Thomson equation,
$\Delta T_\mathrm{m}/T_\mathrm{m,0} = [2 \Delta
\gamma(\infty)]/[\rho_\mathrm{c} \Delta H_\mathrm{m} d_\mathrm{w}]$,
where $\rho_\mathrm{c}$ is the crystalline density, $\Delta H_\mathrm{m}$ the
latent heat of melting,
and $\Delta \gamma(\infty) = \gamma_\mathrm{lw}(\infty) -
\gamma_\mathrm{cw}(\infty)$
the surface tension difference for the crystal/surface and liquid/surface
interfaces.
Considering the values measured from our molecular simulation
($\rho_\mathrm{c} = 37.2\,\mathrm{mol/l}$ and $\Delta H_\mathrm{m} =
27.9\,\mathrm{kJ/mol}$), fitting $\Delta T_\mathrm{m}/T_\mathrm{m,0}$ against
$1/d_\textrm{w}$ leads to $\Delta \gamma(\infty) \sim 0.34 \,\mathrm{J/m^2}$.
To assess the impact of the electrostatic screening length $\lambda$ on
capillary freezing,
we use the following expression in which the liquid/surface and
crystal/surface interfacial
tensions between the metallic surface and the crystal ($x = \mathrm{c}$) or
the liquid ($x = \mathrm{l}$)
is given by its value for the insulating surface corrected for the
charge-image interactions
$U^\textrm{CI}(\lambda)$: $\gamma_{x\mathrm{w}}(\lambda) =
\gamma_{x\mathrm{w}}(\infty)
+ \rho_x \ell U^\textrm{CI}(\lambda)$,
where $\ell$ is a scaling length that converts a volume energy
$U^\textrm{CI}$ into a surface energy.
As shown Supplementary Fig.~1(b), such a simple relationship reasonably
captures the impact of electrostatic screening on the liquid/surface
interfacial tension which was assessed using independent simulation
through the Irving-Kirkwood formalism:
$\gamma(\lambda) = L_z/2 \langle P_\mathrm{N} - P_\mathrm{T}\rangle$,
where the terms in bracket are the average normal and tangential pressures,
$L_z$ is the box length in the $z$ direction and the factor 2 accounts for
the two interfaces in the slit geometry.
Despite the fact that the simple expression $\gamma(\lambda) \sim \rho
U^\textrm{CI}(\lambda)$
neglects the impact of screening on the entropy of the liquid,
it provides an accurate description of the surface tension change
induced by electrostatic screening in the metallic surfaces
(we note that this simple equation holds even better for the crystalline
phase as its entropy is negligible).
This allows writing that
$\Delta \gamma(\lambda) = \Delta \gamma(\infty) + (\rho_\mathrm{l} -
\rho_\mathrm{c})
\ell  U^\textrm{CI}(\lambda)$.
As shown in Fig.\ 5(a), considering that $U^\textrm{CI}(\lambda) < 0$
becomes more negative upon decreasing $\lambda$ and $\rho_\mathrm{c} >
\rho_\mathrm{l}$,
this simple scaling predicts that the shift in the melting point
$\Delta T_\mathrm{m}/T_\mathrm{m,0}$ increases as the surfaces turn from
insulating to metallic.

To confront our results with the experimental data on capillary
freezing,\cite{comtet_2017_nanoscale}
Fig.\ 5(b) shows the impact of the screening length $\lambda$ on the
capillary pore size
$d_\textrm{w}^\mathrm{s}(\lambda) = [2 \Delta \gamma(\lambda)T_\mathrm{m,0}]
/ [\rho_\mathrm{c} \Delta H_\mathrm{m} \Delta T_\mathrm{m}]$
below which salt crystallization is observed.
In more detail, to compare quantitatively our data with those obtained
experimentally for a room temperature ionic liquid, we plot the shift induced
by surface metallicity in this capillary pore size with respect to that for
an insulating surface $\Delta d_\textrm{w}^\mathrm{s}(\lambda)$ where $\Delta
d_\textrm{w}^\mathrm{s}(\lambda) = d_\textrm{w}^\mathrm{s}(\lambda) -
d_\textrm{w}^\mathrm{s}(\infty)$.
The choice to normalize $\Delta d_\textrm{w}^\mathrm{s}(\lambda)$ in Fig.
5(b) by $\Delta d_\textrm{w}^\mathrm{s}(0) = d_\textrm{w}^\mathrm{s}(0) -
d_\textrm{w}^\mathrm{s}(\infty)$ allows defining  a quantity in the $y$ axis
that varies from 0 for a perfect insulator ($\lambda \to \infty$) to 1 for a
perfect metal ($\lambda = 0$). Moreover, in addition to providing a mean to
compare with experimental data for any other system, such a normalized
quantity provides data that are independent of the specifically chosen value
$\Delta T_\mathrm{m}/T_\mathrm{m,0}$. As can be seen in Fig.\ 5(b), our
theoretical predictions do capture the experimentally observed behavior
indicating
that the capillary length $d_\textrm{w}^\mathrm{s}$ increases upon decreasing
the screening
length $\lambda$.
In this plot, the screening length $\lambda$ is normalized by a length $a$,
which is a molecular characteristic of the ionic systems under scrutiny (see
$x$-axis plotted as $a/\lambda$). As shown in Fig. 5(b), a perfect
quantitative agreement between our simulated data for a simple molten salt
and the experimental data for the ionic liquid is observed. This provides the
value $a$ = 1 nm for the simulated molten salt, while a slightly smaller
value $a$ = 0.335 nm was used in Ref.\cite{comtet_2017_nanoscale} in the
analysis of the experimental results for the [Bmim][BF4] room temperature
ionic liquid (though, using the simplified modelling in
Ref.\cite{comtet_2017_nanoscale}). The parameter $a$ can be seen as a
characteristic length describing the impact of electrostatic screening on
freezing. A slightly smaller length a for the room temperature ionic liquids
-- having more complex molecular structure -- suggests a smaller impact of
electrostatic screening on capillary freezing for these complex ions with
respect to a simple salt. This is expected considering that significant
entropy and molecular packing aspects largely affect the crystallization of
room temperature ionic liquids (in particular, these contributions govern
their low melting point). These important results provide a quantitative
microscopic picture for this recent experimental finding in which capillary
freezing of an ionic liquid was found to be promoted by metal surfaces.
Beyond this important result, this study further suggests that the simple
effective approach presented here captures the rich and complex behavior of
charges confined between metallic surfaces.

	\noindent \textbf{Wetting transition} \\
\noindent	Having assessed our effective simulation strategy, we now turn
to the thermodynamically relevant case of the wetting of an ionic
liquid at metal surfaces (as described above, the ionic liquid is taken as
a molten salt modeled using charged particles $\pm 1\rme$).
Fig.\ 6(a) shows
the number density profiles $\rho_n(z)$ for the salt and Thomas--Fermi
fluid for  different $\lambda$. A crossover is observed upon decreasing
$\lambda$; while  the salt is depleted at the insulating interface, a marked
ion
density peak appears under metallic conditions (in contrast, the density
profile for the Thomas--Fermi fluid is nearly unaffected by
$\lambda$). This behavior suggests that the system undergoes a wetting
transition upon changing the dielectric/metallic nature of the
confining medium (perfect wetting/non-wetting for metal/insulator,
respectively).

The observed wetting transition was characterized by measuring the surface
tension of the liquid salt confined at a constant density within surfaces
made of a metallic medium with a screening length  $\lambda$ via the
Irving-Kirkwood formula (introduced above). By considering the salt  in its
liquid ($l$) and gas ($g$) states in contact with the metal ($m$), we
estimated for various $\lambda$ the gas/metal $\gamma_\mathrm{gm}(\lambda)$
and liquid/metal $\gamma_\mathrm{lm}(\lambda)$ surface tensions.
Note that, in molecular dynamics simulations, the various interfaces
(gas/metal, liquid/metal, gas/liquid) are investigated
separately;\cite{nijmeijer_1990_wetting} accordingly, the gas (resp., liquid)
phase is
metastable when the liquid (resp., gas) phase is stable, i.e. wets the
surface. To investigate the impact of surface metallicity on wetting, we then
evaluate the spreading coefficient $S$ from the gas/liquid, surface/liquid
and surface/gas interfacial tensions defined
as\cite{degennes_1985_wetting,rowlinson_1982_molecular} $S =
\gamma_\mathrm{gm} -
\gamma_\mathrm{lm} - \gamma_\mathrm{lg}$. Fig. 6(b) shows the dependence of
the spreading coefficient $S$ on the screening length $\lambda$. This plot
reveals the wetting behavior of the salt solution on the metallic surfaces
under scrutiny, depending on the sign and amplitude of $S$.  As shown in Fig.
6(b),
the sign of the spreading coefficient $S$ changes from $S < 0$ to $S > 0$ for
$\lambda \sim 0.28$ nm, i.e when the nature of the surfaces switches from
insulating to metallic as the screening length $\lambda$ decreases. This is
the signature of a continuous wetting transition of the liquid salt from
partial wetting ($S < 0$) for large $\lambda$ (more insulating surfaces) to
complete wetting ($S > 0$) for small $\lambda$ (more metallic surfaces).
In more detail, for $\lambda <$ 0.28 nm (more metallic surfaces),  $S > 0$
i.e. $\gamma_\mathrm{gm} > \gamma_\mathrm{lm}+ \gamma_\mathrm{lg}$; this
reflects that a wetting film with two interfaces (solid/liquid and
liquid/gas) is of lower surface free energy compared to a solid gas
interface. As a result, in these conditions, the system is perfectly wetting
with a liquid film spreading over the metal surface.
On the other hand, for $\lambda > 0.28$ nm (more insulating surfaces), $S <
0$ i.e. $\gamma_\mathrm{gm} \leq \gamma_\mathrm{lm}+ \gamma_\mathrm{lg}$ so
that the liquid phase wets incompletely the surface. On a macroscopic
surface, this would lead to the formation of a liquid droplet at the solid
surface with a contact angle $\theta$ related to $S$ according to $S =
\gamma_\mathrm{lg} (\cos \theta -
1)$.\cite{degennes_1985_wetting,rowlinson_1982_molecular}

The data in Fig.\ 6(b) for partial wetting suggest that $\cos\theta$ tends to
1 in a linear fashion upon decreasing the screening length $\lambda$. As
discussed in Ref.\cite{bonn_2009_wetting}, this scaling suggests that the
wetting transition induced by tuning the solid surface from an imperfect to
perfect metal is a first-order transition.
Moreover, increasing the screening length $\lambda$ beyond 1 nm (more and
more insulating surface) is expected to lead to complete drying ($\cos \theta
=-1$).
As discussed in Ref.\cite{evans_2019_unified} for simple liquids, such drying
transition is expected to be a second-order transition in contrast to the
wetting transition discussed above.
As shown in the inset of Fig.\ 6(b), the change in $\Delta
\gamma$ between the insulator and metal is found to scale with the
liquid/gas density contrast:
\begin{equation}
    \Delta(\Delta \gamma(\lambda)) = \Delta \gamma(\lambda) - \Delta
    \gamma(\infty)
    \sim (\rho_\mathrm{l} - \rho_\mathrm{g})
    \alpha(\lambda) \sim \rho_\mathrm{l} \alpha(\lambda)
    \label{eq:shift_rho}
\end{equation}
where $\rho_\mathrm{l} \gg \rho_\mathrm{g}$ was assumed in the second
equality.
As expected from the Thomas--Fermi model, the inset in Fig.\ 6(b)
shows that $\alpha(\lambda) \sim U_{\lambda}^{\mathrm{CI}}$ as the charge
interaction with the induced density distributions (including the charge
image) is dominating the surface energy excess.

Despite the key role of electrostatic interactions --- including screening
induced by metallic surfaces --- on the behavior of charges near surfaces, we
emphasize that surface wetting is also strongly affected by the so-called
ion-specific effects. Like in bulk electrolytes, these effects which arise
from the ion molecular structure give rise to a complex physicochemical
behavior interaction of charges and dipolar molecules near the surface.
Noteworthy, the classical Frumkin-Damaskin theory describes the relative
strength of electrostatic interactions in the vicinity of a charged electrode
with respect to interactions responsible for the adsorption of small polar
molecules. This model leads to the so-called Frumkin adsorption isotherm
which describes how an electrode polarization increase induces desorption of
polar molecules concomitantly with the adsorption of water and
ions.\cite{damaskin_2012_adsorption} In this context, owing to its
versatility, our molecular strategy of electrostatic screening between
metallic surfaces is suited to account for such molecular and physicochemical
effects since it relies on a general molecular dynamics approach that can be
employed with any available force field. In fact, this is one of the assets
of this effective approach that it can be used for ionic systems (regardless
of the ion structure complexity) but also dipolar liquids which are expected
to be affected by electrostatic screening when confined between metallic
surfaces.

	\begin{figure}[htbp]
    \begin{center}
        \includegraphics[width=.6\columnwidth]{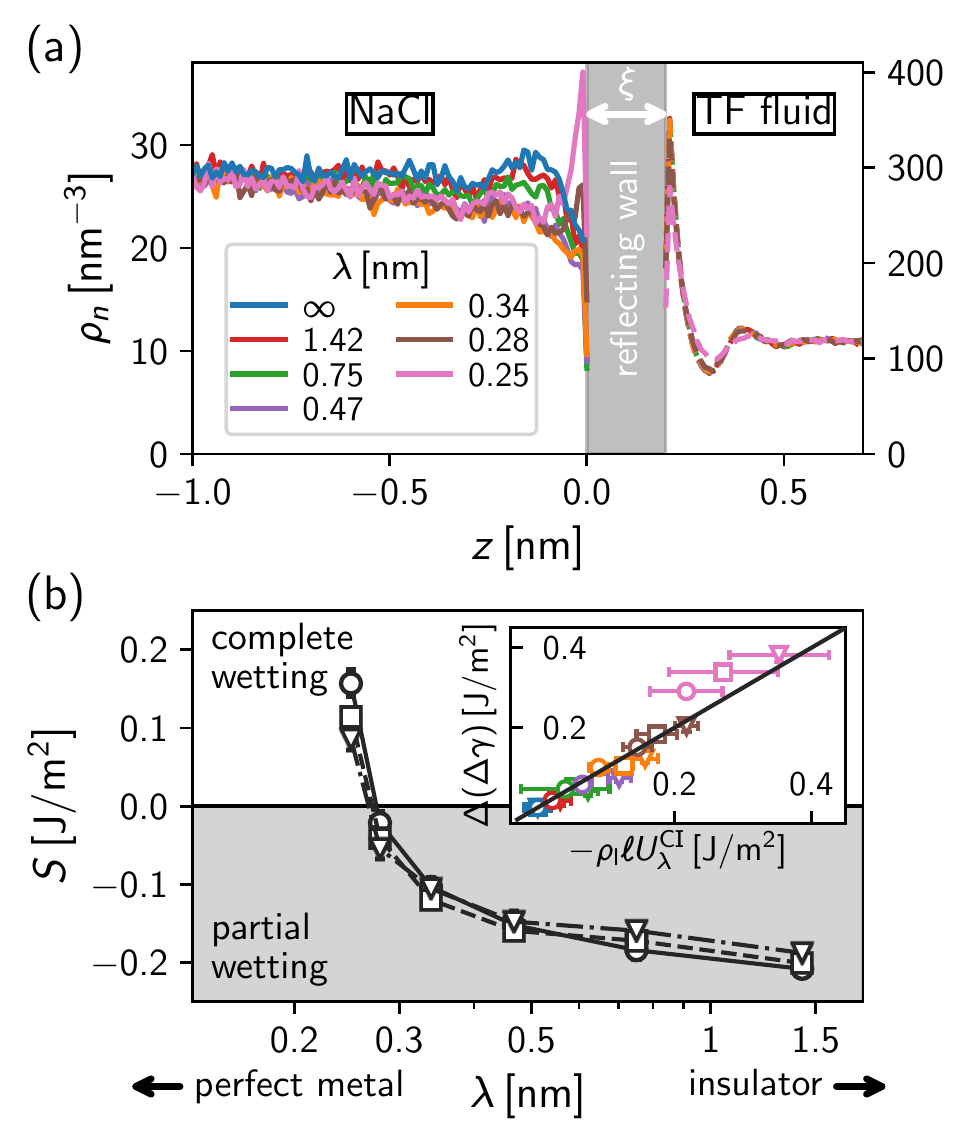}
    \end{center}
    \caption{\textbf{Wetting transition of ionic liquids at metal surfaces.}
        (a) Number density profile $\rho_n$ for a molten salt confined
        between metal
        surfaces with different screening lengths $\lambda$. The
        Thomas--Fermi fluid within the metal (right) and the
        molten salt (left) are separated by a reflective wall of
        thickness $\xi = 0.2$ nm.
        The average salt density is  $\rho_\mathrm{l} =
        27\,\mathrm{nm}^{-3}$.
        (b) Surface tension difference $\Delta \gamma =
        \gamma_\mathrm{gm}-\gamma_\mathrm{lm}$ normalized by the gas/liquid
        surface tension $\gamma_\mathrm{gl}$ as a function of $\lambda$.
        Symbols indicate different liquid densities $\rho_\mathrm{l}$:
        $27\,\mathrm{nm}^{-3}$ (circles), $25\,\mathrm{nm}^{-3}$
        (squares) and $23\,\mathrm{nm}^{-3}$ (triangles).
        The inset shows the change in the surface tension difference
        $\Delta(\Delta \gamma(\lambda))$ between an insulator and
        the Thomas--Fermi fluid at a given $\lambda$.
        The characteristic length $\ell$ converts the volume energy to a
        surface energy.
        Symbols correspond to the same densities as in the main figure,
        color coding denotes different $\lambda$ as in (a).
    }
    \label{fig:fig_6}
\end{figure}

	\noindent \textbf{Discussion} \\
\noindent		We developed a classical molecular simulation strategy that
allows considering the confinement within any material ranging from perfect
metal to insulator. This approach, which does not require to input any
given geometry/molecular structure for the confining material, describes in
an effective fashion electrostatic screening within confined/vicinal
fluids together with the expected capacitive behavior.
After straightforward integration into existing simulation
packages, this method offers a useful framework to investigate the behavior
of dipolar and charged fluids in porous materials made up of any material
with imperfect dielectric/metal properties. Beyond practical implications,
we also unraveled a non-wetting/wetting crossover in nanoconfined liquids
as the confining surfaces vary from insulator to perfect metal. This raises
new challenging questions on the complex behavior of charged systems in the
vicinity or confined within surfaces with important applications such as
electrowetting/switching for energy storage, lubrication, catalysis, etc.

\section*{Acknowledgements}
We acknowledge V. Kaiser for his help with the Thomas--Fermi
model and computation time through CIMENT infrastructure (Rhône-Alpes
CPER07\_13 CIRA) and Equip@Meso project (ANR-10-EQPX-29-01). We also
acknowledge funding from the ANR project TAMTAM (ANR-15-CE08-0008-01).
AS acknowledges funding from the DFG under Germany's Excellence Strategy ---
EXC 2075 --- 390740016 and SFB 1313 (Project Number 327154368) and support by
the
Stuttgart Center for Simulation Science (SimTech).

\section*{Author contributions}
B.C.\, L.B.\ and A.S.\ conceived the
research. A.S.\ carried out the molecular simulations with support from
D.J.
A.S., B.C.\ and L.B.\ analyzed the data. A.S.\ and B.C.\ wrote the
paper with inputs from all authors.

\section*{Competing Interests}
The authors declare no competing interests.

	\begin{methods}

    \subsection{Molecular Dynamics simulations.}
    All simulations are carried out using LAMMPS simulation
    package\cite{plimpton_1995_fast} (stable release 7 Aug 2019).
    Electrostatic interactions are calculated using the PPPM
    method with an accuracy of at least
    $10^{-4}$ and a real-space cut-off $r_\mathrm{c}=12.5\,\textrm{\AA}$.
    Periodic boundary conditions are used in all dimensions with
    the non-electrostatic interactions being cut and shifted to zero at
    $r_\mathrm{c}$.
    For the simulations of the TF fluid and the salt in contact with an
    insulating/vacuum interface, the interactions between periodic images
    are not screened so that we employ the slab correction method proposed
    by Yeh and
    Berkowitz\cite{yeh_1999_ewald} with a vacuum layer of three times the
    simulation cell height.
    The non-electrostatic part of the salt--salt interactions are
    described using the Born--Meyer--Huggins
    potential which accurately reproduces the properties of  NaCl (either
    as a crystal or molten salt),\cite{anwar_2002_calculation}
    \begin{equation}
        U_\mathrm{BMH}(r) = A \exp\left(\frac{\sigma-r}{B}\right) -
        \frac{C}{r^6} -
        \frac{D}{r^8},
    \end{equation}
    The corresponding force field parameters are
    given in \textit{Supplementary Table 1}.
    Reflective walls of width $\xi=0.2\,\mathrm{nm}$ are used at each
    metal/dielectric
    interface to prevent the Thomas--Fermi fluid/charged system to migrate
    to the pore space/confining media.
    The latter implies that, if an atom moves through the wall by a
    distance $\delta$ in a timestep, its position is set back to $-\delta$
    away from the wall and the sign of the corresponding velocity component
    is flipped.

    In all simulations presented in the main text, the confining
    media filled with the Thomas--Fermi fluid are chosen to have a length
    $d_\mathrm{TF} = 10\,\mathrm{nm}$. Increasing $d_\mathrm{TF}$ increases
    the agreement between theory and simulations in Fig.\ 3 due to
    the	decay in the disjoining energy between the two TF surfaces but at
    the price of enhanced numerical cost (\textit{Supplementary Fig.~S10}).
    For the TF--TF interaction, a purely repulsive power
    law of the form	$U(r) = E/r^n$ is added to avoid numerical infinities
    when particles overlap.
    We use $n=8$ and $E=10^3 \,\textrm{kcal/mol/\AA}^8$ but we checked that
    the	detailed form of the interaction potential does not qualitatively
    influence our simulation results as shown in \textit{Supplementary
        Fig.~S9}.
    The positive and negative TF particles differ only in their
    partial charge $\pm q_\mathrm{TF}$ and only interact through
    electrostatic interactions
    with the salt.
    The density of the TF fluid is fixed at $\rho_\mathrm{TF} =
    57.5\,\mathrm{nm^{-3}}$
    at a temperature $T_\mathrm{TF} = 12000\,\mathrm{K}$ and mass
    $m_\mathrm{TF} = 1 \,\mathrm{amu}$ to ensure fast relaxation.
    The mass of the Na and Cl atoms is set to 22.9898 and 35.446 amu,
    respectively.
    Time integration is performed using a Verlet scheme with a timestep of
    $0.1\,\mathrm{fs}$ to allow for fast relaxation of the TF liquid.
    The molten salt is simulated at a 2000~K and temperature coupling
    is performed using separate Nose--Hoover thermostats for the salt and
    TF fluids with a characteristic time of 100 timesteps.

    \subsection{Capacitance determination.}
    The capacitive behavior of our virtual Thomas--Fermi fluid was checked
    as this provides an important benchmark to assess its physical
    validity. Using a direct measurement approach, the capacitance was
    estimated using MD simulations in which the system is sandwiched
    between two electrodes having an overall charge $+Q$ and $-Q$. As
    discussed in the main text, two systems were considered to verify the
    consistency of the obtained results: the virtual Thomas--Fermi alone
    and a composite system made up of a dielectric layer confined by the
    virtual Thomas--Fermi fluid (for the latter,  two dielectric materials
    were considered: either a vacuum layer or a molten salt). The
    electrodes used for the capacitance measurements consist of point
    charges $q_\mathrm{w} = 0.01$ arranged on a 1\AA\, 2D grid (see
    \textit{Supplementary Fig.~S2} for a molecular simulation snapshot),
    resulting in
    a total charge of $Q=\pm 0.166\,\mathrm{C/m^2}$. It was  checked that
    this value is low enough to ensure that the capacitance response of the
    system is in the  linear response regime so that the capcacitance $C$
    is readily obtained from the electrostatic potential drop $\Delta
    \Psi$ between the two electrodes.
    The TF fluid is separated from the point charges by
    1\AA\ via a reflecting wall denoted by the gray shaded areas in
    Fig.\ 4(a).
    The potential drop is obtained from Poisson equation by
    integrating twice the charge density profile, $\Delta \Psi(z) =
    - \int_{-\infty}^z \rmd z^\prime \int_{-\infty}^{z\prime} \rmd
    z^{\prime\prime} \,e(\rho_+ - \rho_-)/\varepsilon_0$ as
    shown in \textit{Supplementary Fig.~S2(b)}.

    \section*{Data availability}
    All relevant simulation input scripts are available in this repository:
    Schlaich, Alexander, 2021, "Simulation input scripts for 'Electronic
    screening using a virtual Thomas–Fermi fluid for predicting wetting and
    phase transitions of ionic liquids at metal surfaces'",
    https://doi.org/10.18419/darus-2115, DaRUS.

    \section*{Code availability}
    Molecular simulations were performed using using the open source package
    LAMMPS, stable release 7 Aug 2019, available under
    {https://www.lammps.org/}.
    Post-processing has been performed in Python using our open source
    toolbox MAICoS ({https://gitlab.com/maicos-devel/maicos/}).

\end{methods}

\bibliography{TF}

\begin{thebibliography}{10}
\expandafter\ifx\csname url\endcsname\relax
  \def\url#1{\texttt{#1}}\fi
\expandafter\ifx\csname urlprefix\endcsname\relax\def\urlprefix{URL }\fi
\providecommand{\bibinfo}[2]{#2}
\providecommand{\eprint}[2][]{\url{#2}}

\bibitem{bocquet_2010_nanofluidics}
\bibinfo{author}{Bocquet, L.} \& \bibinfo{author}{Charlaix, E.}
\newblock \bibinfo{title}{Nanofluidics, from bulk to interfaces}.
\newblock \emph{\bibinfo{journal}{Chem. Soc. Rev.}}
  \textbf{\bibinfo{volume}{39}}, \bibinfo{pages}{1073--1095}
  (\bibinfo{year}{2010}).

\bibitem{schoch_2008_transport}
\bibinfo{author}{Schoch, R.~B.}, \bibinfo{author}{Han, J.} \&
  \bibinfo{author}{Renaud, P.}
\newblock \bibinfo{title}{Transport phenomena in nanofluidics}.
\newblock \emph{\bibinfo{journal}{Rev. Mod. Phys.}}
  \textbf{\bibinfo{volume}{80}}, \bibinfo{pages}{839--883}
  (\bibinfo{year}{2008}).

\bibitem{bazant_2011_double}
\bibinfo{author}{Bazant, M.~Z.}, \bibinfo{author}{Storey, B.~D.} \&
  \bibinfo{author}{Kornyshev, A.~A.}
\newblock \bibinfo{title}{Double {{Layer}} in {{Ionic Liquids}}: Overscreening
  versus {{Crowding}}}.
\newblock \emph{\bibinfo{journal}{Phys. Rev. Lett.}}
  \textbf{\bibinfo{volume}{106}}, \bibinfo{pages}{046102}
  (\bibinfo{year}{2011}).

\bibitem{smith_2016_electrostatic}
\bibinfo{author}{Smith, A.~M.}, \bibinfo{author}{Lee, A.~A.} \&
  \bibinfo{author}{Perkin, S.}
\newblock \bibinfo{title}{The {{Electrostatic Screening Length}} in
  {{Concentrated Electrolytes Increases}} with {{Concentration}}}.
\newblock \emph{\bibinfo{journal}{J. Phys. Chem. Lett.}}
  \textbf{\bibinfo{volume}{7}}, \bibinfo{pages}{2157--2163}
  (\bibinfo{year}{2016}).

\bibitem{laine_2020_nanotribology}
\bibinfo{author}{Lain{\'e}, A.}, \bibinfo{author}{Nigu{\`e}s, A.},
  \bibinfo{author}{Bocquet, L.} \& \bibinfo{author}{Siria, A.}
\newblock \bibinfo{title}{Nanotribology of {{Ionic Liquids}}: Transition to
  {{Yielding Response}} in {{Nanometric Confinement}} with {{Metallic
  Surfaces}}}.
\newblock \emph{\bibinfo{journal}{Phys. Rev. X}} \textbf{\bibinfo{volume}{10}},
  \bibinfo{pages}{011068} (\bibinfo{year}{2020}).

\bibitem{fedorov_2014_ionic}
\bibinfo{author}{Fedorov, M.~V.} \& \bibinfo{author}{Kornyshev, A.~A.}
\newblock \bibinfo{title}{Ionic {{Liquids}} at {{Electrified Interfaces}}}.
\newblock \emph{\bibinfo{journal}{Chem. Rev.}} \textbf{\bibinfo{volume}{114}},
  \bibinfo{pages}{2978--3036} (\bibinfo{year}{2014}).

\bibitem{kaiser_2017_electrostatic}
\bibinfo{author}{Kaiser, V.} \emph{et~al.}
\newblock \bibinfo{title}{Electrostatic interactions between ions near
  {{Thomas}}\textendash{{Fermi}} substrates and the surface energy of ionic
  crystals at imperfect metals}.
\newblock \emph{\bibinfo{journal}{Faraday Discuss.}}
  \textbf{\bibinfo{volume}{199}}, \bibinfo{pages}{129--158}
  (\bibinfo{year}{2017}).

\bibitem{dufils_2019_semiclassical}
\bibinfo{author}{Dufils, T.}, \bibinfo{author}{Scalfi, L.},
  \bibinfo{author}{Rotenberg, B.} \& \bibinfo{author}{Salanne, M.}
\newblock \bibinfo{title}{A semiclassical {{Thomas}}-{{Fermi}} model to tune
  the metallicity of electrodes in molecular simulations}.
\newblock \emph{\bibinfo{journal}{arXiv:1910.13341 [cond-mat]}}
  (\bibinfo{year}{2019}).
\newblock \eprint{1910.13341}.

\bibitem{newns_1969_fermi}
\bibinfo{author}{Newns, D.~M.}
\newblock \bibinfo{title}{Fermi\textendash{{Thomas Response}} of a {{Metal
  Surface}} to an {{External Point Charge}}}.
\newblock \emph{\bibinfo{journal}{J. Chem. Phys.}}
  \textbf{\bibinfo{volume}{50}}, \bibinfo{pages}{4572--4575}
  (\bibinfo{year}{1969}).

\bibitem{inkson_1973_manybody}
\bibinfo{author}{Inkson, J.~C.}
\newblock \bibinfo{title}{Many-body effect at metal-semiconductor junctions.
  {{II}}. {{The}} self energy and band structure distortion}.
\newblock \emph{\bibinfo{journal}{J. Phys. C: Solid State Phys.}}
  \textbf{\bibinfo{volume}{6}}, \bibinfo{pages}{1350--1362}
  (\bibinfo{year}{1973}).

\bibitem{kornyshev_1977_image}
\bibinfo{author}{Kornyshev, A.~A.}, \bibinfo{author}{Rubinshtein, A.~I.} \&
  \bibinfo{author}{Vorotyntsev, M.~A.}
\newblock \bibinfo{title}{Image potential near a dielectric\textendash
  plasma-like medium interface}.
\newblock \emph{\bibinfo{journal}{physica status solidi (b)}}
  \textbf{\bibinfo{volume}{84}}, \bibinfo{pages}{125--132}
  (\bibinfo{year}{1977}).

\bibitem{luque_2012_electric}
\bibinfo{author}{Luque, N.~B.} \& \bibinfo{author}{Schmickler, W.}
\newblock \bibinfo{title}{The electric double layer on graphite}.
\newblock \emph{\bibinfo{journal}{Electrochimica Acta}}
  \textbf{\bibinfo{volume}{71}}, \bibinfo{pages}{82--85}
  (\bibinfo{year}{2012}).

\bibitem{kornyshev_2014_differential}
\bibinfo{author}{Kornyshev, A.~A.}, \bibinfo{author}{Luque, N.~B.} \&
  \bibinfo{author}{Schmickler, W.}
\newblock \bibinfo{title}{Differential capacitance of ionic liquid interface
  with graphite: The story of two double layers}.
\newblock \emph{\bibinfo{journal}{J Solid State Electrochem}}
  \textbf{\bibinfo{volume}{18}}, \bibinfo{pages}{1345--1349}
  (\bibinfo{year}{2014}).

\bibitem{netz_1999_debyehuckel}
\bibinfo{author}{Netz, R.~R.}
\newblock \bibinfo{title}{Debye-{{H\"uckel}} theory for interfacial
  geometries}.
\newblock \emph{\bibinfo{journal}{Phys. Rev. E}} \textbf{\bibinfo{volume}{60}},
  \bibinfo{pages}{3174--3182} (\bibinfo{year}{1999}).

\bibitem{lee_2016_ion}
\bibinfo{author}{Lee, A.~A.} \& \bibinfo{author}{Perkin, S.}
\newblock \bibinfo{title}{Ion\textendash{{Image Interactions}} and {{Phase
  Transition}} at {{Electrolyte}}\textendash{{Metal Interfaces}}}.
\newblock \emph{\bibinfo{journal}{J. Phys. Chem. Lett.}}
  \textbf{\bibinfo{volume}{7}}, \bibinfo{pages}{2753--2757}
  (\bibinfo{year}{2016}).

\bibitem{bedrov_2019_molecular}
\bibinfo{author}{Bedrov, D.} \emph{et~al.}
\newblock \bibinfo{title}{Molecular {{Dynamics Simulations}} of {{Ionic
  Liquids}} and {{Electrolytes Using Polarizable Force Fields}}}.
\newblock \emph{\bibinfo{journal}{Chem. Rev.}} \textbf{\bibinfo{volume}{119}},
  \bibinfo{pages}{7940--7995} (\bibinfo{year}{2019}).

\bibitem{breitsprecher_2015_electrode}
\bibinfo{author}{Breitsprecher, K.}, \bibinfo{author}{Szuttor, K.} \&
  \bibinfo{author}{Holm, C.}
\newblock \bibinfo{title}{Electrode {{Models}} for {{Ionic Liquid}}-{{Based
  Capacitors}}}.
\newblock \emph{\bibinfo{journal}{J. Phys. Chem. C}}
  \textbf{\bibinfo{volume}{119}}, \bibinfo{pages}{22445--22451}
  (\bibinfo{year}{2015}).

\bibitem{comtet_2017_nanoscale}
\bibinfo{author}{Comtet, J.} \emph{et~al.}
\newblock \bibinfo{title}{Nanoscale capillary freezing of ionic liquids
  confined between metallic interfaces and the role of electronic screening}.
\newblock \emph{\bibinfo{journal}{Nature Materials}}
  \textbf{\bibinfo{volume}{16}}, \bibinfo{pages}{634--639}
  (\bibinfo{year}{2017}).

\bibitem{ashcroft_1976_solid}
\bibinfo{author}{Ashcroft, N.~W.} \& \bibinfo{author}{Mermin, N.~D.}
\newblock \emph{\bibinfo{title}{Solid {{State Physics}}}}
  (\bibinfo{publisher}{{Holt, Rinehart and Winston}}, \bibinfo{year}{1976}).

\bibitem{dossantos_2017_simulations}
\bibinfo{author}{{dos Santos}, A.~P.}, \bibinfo{author}{Girotto, M.} \&
  \bibinfo{author}{Levin, Y.}
\newblock \bibinfo{title}{Simulations of {{Coulomb}} systems confined by
  polarizable surfaces using periodic {{Green}} functions}.
\newblock \emph{\bibinfo{journal}{J. Chem. Phys.}}
  \textbf{\bibinfo{volume}{147}}, \bibinfo{pages}{184105}
  (\bibinfo{year}{2017}).

\bibitem{siepmann_1995_influence}
\bibinfo{author}{Siepmann, J.~I.} \& \bibinfo{author}{Sprik, M.}
\newblock \bibinfo{title}{Influence of surface topology and electrostatic
  potential on water/electrode systems}.
\newblock \emph{\bibinfo{journal}{J. Chem. Phys.}}
  \textbf{\bibinfo{volume}{102}}, \bibinfo{pages}{511--524}
  (\bibinfo{year}{1995}).

\bibitem{reed_2007_electrochemical}
\bibinfo{author}{Reed, S.~K.}, \bibinfo{author}{Lanning, O.~J.} \&
  \bibinfo{author}{Madden, P.~A.}
\newblock \bibinfo{title}{Electrochemical interface between an ionic liquid and
  a model metallic electrode}.
\newblock \emph{\bibinfo{journal}{The Journal of Chemical Physics}}
  \textbf{\bibinfo{volume}{126}}, \bibinfo{pages}{084704}
  (\bibinfo{year}{2007}).

\bibitem{tyagi_2010_iterative}
\bibinfo{author}{Tyagi, S.} \emph{et~al.}
\newblock \bibinfo{title}{An iterative, fast, linear-scaling method for
  computing induced charges on arbitrary dielectric boundaries}.
\newblock \emph{\bibinfo{journal}{J. Chem. Phys.}}
  \textbf{\bibinfo{volume}{132}}, \bibinfo{pages}{154112}
  (\bibinfo{year}{2010}).

\bibitem{arnold_2013_efficient}
\bibinfo{author}{Arnold, A.} \emph{et~al.}
\newblock \bibinfo{title}{Efficient {{Algorithms}} for {{Electrostatic
  Interactions Including Dielectric Contrasts}}}.
\newblock \emph{\bibinfo{journal}{Entropy}} \textbf{\bibinfo{volume}{15}},
  \bibinfo{pages}{4569--4588} (\bibinfo{year}{2013}).

\bibitem{nguyen_2019_incorporating}
\bibinfo{author}{Nguyen, T.~D.}, \bibinfo{author}{Li, H.},
  \bibinfo{author}{Bagchi, D.}, \bibinfo{author}{Solis, F.~J.} \&
  \bibinfo{author}{{Olvera de la Cruz}, M.}
\newblock \bibinfo{title}{Incorporating surface polarization effects into
  large-scale coarse-grained {{Molecular Dynamics}} simulation}.
\newblock \emph{\bibinfo{journal}{Computer Physics Communications}}
  \textbf{\bibinfo{volume}{241}}, \bibinfo{pages}{80--91}
  (\bibinfo{year}{2019}).

\bibitem{torrie_1986_double}
\bibinfo{author}{Torrie, G.~M.} \& \bibinfo{author}{Valleau, J.~P.}
\newblock \bibinfo{title}{Double layer structure at the interface between two
  immiscible electrolyte solutions}.
\newblock \emph{\bibinfo{journal}{Journal of Electroanalytical Chemistry and
  Interfacial Electrochemistry}} \textbf{\bibinfo{volume}{206}},
  \bibinfo{pages}{69--79} (\bibinfo{year}{1986}).

\bibitem{kornyshev_1980_nonlocal}
\bibinfo{author}{Kornyshev, A.~A.} \& \bibinfo{author}{Vorotyntsev, M.~A.}
\newblock \bibinfo{title}{Nonlocal electrostatic approach to the double layer
  and adsorption at the electrode-electrolyte interface}.
\newblock \emph{\bibinfo{journal}{Surface Science}}
  \textbf{\bibinfo{volume}{101}}, \bibinfo{pages}{23--48}
  (\bibinfo{year}{1980}).

\bibitem{vorotyntsev_1986_modern}
\bibinfo{author}{Vorotyntsev, M.~A.}
\newblock \bibinfo{title}{Modern {{State}} of {{Double Layer Study}} of {{Solid
  Metals}}}.
\newblock In \bibinfo{editor}{Bockris, J.~O.}, \bibinfo{editor}{Conway, B.~E.}
  \& \bibinfo{editor}{White, R.~E.} (eds.) \emph{\bibinfo{booktitle}{Modern
  {{Aspects}} of {{Electrochemistry}}: Volume 17}}, Modern {{Aspects}} of
  {{Electrochemistry}}, \bibinfo{pages}{131--222}
  (\bibinfo{publisher}{{Springer US}}, \bibinfo{address}{{Boston, MA}},
  \bibinfo{year}{1986}).

\bibitem{kornyshev_1980_electrostatic}
\bibinfo{author}{Kornyshev, A.~A.} \& \bibinfo{author}{Vorotyntsev, M.~A.}
\newblock \bibinfo{title}{Electrostatic {{Interaction}} at the metal/dielectric
  interface}.
\newblock \emph{\bibinfo{journal}{Sov. Phys. JETP}}
  \textbf{\bibinfo{volume}{51}}, \bibinfo{pages}{509--513}
  (\bibinfo{year}{1980}).

\bibitem{vorotyntsev_1988_}
\bibinfo{author}{Vorotyntsev, M.~A.}
\newblock No. \bibinfo{number}{Part C} in \bibinfo{series}{The Chemical Physics
  of Solvation}, \bibinfo{pages}{401} (\bibinfo{publisher}{{Elsevier}},
  \bibinfo{address}{{Amsterdam}}, \bibinfo{year}{1988}).

\bibitem{kornyshev_1986_coverage}
\bibinfo{author}{Kornyshev, A.~A.} \& \bibinfo{author}{Schmickler, W.}
\newblock \bibinfo{title}{On the coverage dependence of the partial charge
  transfer coefficient}.
\newblock \emph{\bibinfo{journal}{Journal of Electroanalytical Chemistry and
  Interfacial Electrochemistry}} \textbf{\bibinfo{volume}{202}},
  \bibinfo{pages}{1--21} (\bibinfo{year}{1986}).

\bibitem{vorotyntsev_1980_possible}
\bibinfo{author}{Vorotyntsev, M.}, \bibinfo{author}{Kornyshev, A.} \&
  \bibinfo{author}{Rubinshtein, A.}
\newblock \bibinfo{title}{Possible {{Mechanisms}} of {{Controlling Ionic
  Interaction}} at the {{Electrode}}-{{Solution Interface}}}.
\newblock \emph{\bibinfo{journal}{Soviet Electrochemistry}}
  \textbf{\bibinfo{volume}{16}}, \bibinfo{pages}{65--67}
  (\bibinfo{year}{1980}).

\bibitem{kornyshev_1989_metal}
\bibinfo{author}{Kornyshev, A.~A.}
\newblock \bibinfo{title}{Metal electrons in the double layer theory}.
\newblock \emph{\bibinfo{journal}{Electrochimica Acta}}
  \textbf{\bibinfo{volume}{34}}, \bibinfo{pages}{1829--1847}
  (\bibinfo{year}{1989}).

\bibitem{gerischer_1985_interpretation}
\bibinfo{author}{Gerischer, H.}
\newblock \bibinfo{title}{An interpretation of the double layer capacity of
  graphite electrodes in relation to the density of states at the {{Fermi}}
  level}.
\newblock \emph{\bibinfo{journal}{J. Phys. Chem.}}
  \textbf{\bibinfo{volume}{89}}, \bibinfo{pages}{4249--4251}
  (\bibinfo{year}{1985}).

\bibitem{gerischer_1987_density}
\bibinfo{author}{Gerischer, H.}, \bibinfo{author}{McIntyre, R.},
  \bibinfo{author}{Scherson, D.} \& \bibinfo{author}{Storck, W.}
\newblock \bibinfo{title}{Density of the electronic states of graphite:
  Derivation from differential capacitance measurements}.
\newblock \emph{\bibinfo{journal}{J. Phys. Chem.}}
  \textbf{\bibinfo{volume}{91}}, \bibinfo{pages}{1930--1935}
  (\bibinfo{year}{1987}).

\bibitem{kondrat_2010_superionic}
\bibinfo{author}{Kondrat, S.} \& \bibinfo{author}{Kornyshev, A.}
\newblock \bibinfo{title}{Superionic state in double-layer capacitors with
  nanoporous electrodes}.
\newblock \emph{\bibinfo{journal}{J. Phys.: Condens. Matter}}
  \textbf{\bibinfo{volume}{23}}, \bibinfo{pages}{022201}
  (\bibinfo{year}{2010}).

\bibitem{li_2018_computer}
\bibinfo{author}{Li, Z.}, \bibinfo{author}{{Mendez-Morales}, T.} \&
  \bibinfo{author}{Salanne, M.}
\newblock \bibinfo{title}{Computer simulation studies of nanoporous
  carbon-based electrochemical capacitors}.
\newblock \emph{\bibinfo{journal}{Current Opinion in Electrochemistry}}
  \textbf{\bibinfo{volume}{9}}, \bibinfo{pages}{81--86} (\bibinfo{year}{2018}).

\bibitem{rochester_2013_interionic}
\bibinfo{author}{Rochester, C.~C.}, \bibinfo{author}{Lee, A.~A.},
  \bibinfo{author}{Pruessner, G.} \& \bibinfo{author}{Kornyshev, A.~A.}
\newblock \bibinfo{title}{Interionic {{Interactions}} in {{Conducting
  Nanoconfinement}}}.
\newblock \emph{\bibinfo{journal}{ChemPhysChem}} \textbf{\bibinfo{volume}{14}},
  \bibinfo{pages}{4121--4125} (\bibinfo{year}{2013}).

\bibitem{mohammadzadeh_2016_energetics}
\bibinfo{author}{Mohammadzadeh, L.} \emph{et~al.}
\newblock \bibinfo{title}{On the {{Energetics}} of {{Ions}} in {{Carbon}} and
  {{Gold Nanotubes}}}.
\newblock \emph{\bibinfo{journal}{ChemPhysChem}} \textbf{\bibinfo{volume}{17}},
  \bibinfo{pages}{78--85} (\bibinfo{year}{2016}).

\bibitem{bi_2018_minimizing}
\bibinfo{author}{Bi, S.} \emph{et~al.}
\newblock \bibinfo{title}{Minimizing the electrosorption of water from humid
  ionic liquids on electrodes}.
\newblock \emph{\bibinfo{journal}{Nature Communications}}
  \textbf{\bibinfo{volume}{9}}, \bibinfo{pages}{5222} (\bibinfo{year}{2018}).

\bibitem{anwar_2002_calculation}
\bibinfo{author}{Anwar, J.}, \bibinfo{author}{Frenkel, D.} \&
  \bibinfo{author}{Noro, M.~G.}
\newblock \bibinfo{title}{Calculation of the melting point of {{NaCl}} by
  molecular simulation}.
\newblock \emph{\bibinfo{journal}{J. Chem. Phys.}}
  \textbf{\bibinfo{volume}{118}}, \bibinfo{pages}{728--735}
  (\bibinfo{year}{2002}).

\bibitem{nijmeijer_1990_wetting}
\bibinfo{author}{Nijmeijer, M. J.~P.}, \bibinfo{author}{Bruin, C.},
  \bibinfo{author}{Bakker, A.~F.} \& \bibinfo{author}{{van Leeuwen}, J. M.~J.}
\newblock \bibinfo{title}{Wetting and drying of an inert wall by a fluid in a
  molecular-dynamics simulation}.
\newblock \emph{\bibinfo{journal}{Phys. Rev. A}} \textbf{\bibinfo{volume}{42}},
  \bibinfo{pages}{6052--6059} (\bibinfo{year}{1990}).

\bibitem{degennes_1985_wetting}
\bibinfo{author}{{de Gennes}, P.~G.}
\newblock \bibinfo{title}{Wetting: Statics and dynamics}.
\newblock \emph{\bibinfo{journal}{Rev. Mod. Phys.}}
  \textbf{\bibinfo{volume}{57}}, \bibinfo{pages}{827--863}
  (\bibinfo{year}{1985}).

\bibitem{rowlinson_1982_molecular}
\bibinfo{author}{Rowlinson, J.~S.} \& \bibinfo{author}{Widom, B.}
\newblock \emph{\bibinfo{title}{Molecular Theory of Capillarity}}
  (\bibinfo{publisher}{{Courier Corporation}}, \bibinfo{year}{1982}).

\bibitem{bonn_2009_wetting}
\bibinfo{author}{Bonn, D.}, \bibinfo{author}{Eggers, J.},
  \bibinfo{author}{Indekeu, J.}, \bibinfo{author}{Meunier, J.} \&
  \bibinfo{author}{Rolley, E.}
\newblock \bibinfo{title}{Wetting and spreading}.
\newblock \emph{\bibinfo{journal}{Rev. Mod. Phys.}}
  \textbf{\bibinfo{volume}{81}}, \bibinfo{pages}{739--805}
  (\bibinfo{year}{2009}).

\bibitem{evans_2019_unified}
\bibinfo{author}{Evans, R.}, \bibinfo{author}{Stewart, M.~C.} \&
  \bibinfo{author}{Wilding, N.~B.}
\newblock \bibinfo{title}{A unified description of hydrophilic and
  superhydrophobic surfaces in terms of the wetting and drying transitions of
  liquids}.
\newblock \emph{\bibinfo{journal}{PNAS}} \textbf{\bibinfo{volume}{116}},
  \bibinfo{pages}{23901--23908} (\bibinfo{year}{2019}).

\bibitem{damaskin_2012_adsorption}
\bibinfo{author}{Damaskin, B.}
\newblock \emph{\bibinfo{title}{{Adsorption of Organic Compounds on
  Electrodes}}} (\bibinfo{publisher}{{Springer}}, \bibinfo{address}{{Boston,
  MA}}, \bibinfo{year}{2012}), \bibinfo{edition}{softcover reprint of the
  original 1st ed. 1971 edition} edn.

\bibitem{plimpton_1995_fast}
\bibinfo{author}{Plimpton, S.}
\newblock \bibinfo{title}{Fast {{Parallel Algorithms}} for {{Short}}-{{Range
  Molecular Dynamics}}}.
\newblock \emph{\bibinfo{journal}{J. Comput. Phys.}}
  \textbf{\bibinfo{volume}{117}}, \bibinfo{pages}{1--19}
  (\bibinfo{year}{1995}).

\bibitem{yeh_1999_ewald}
\bibinfo{author}{Yeh, I.-C.} \& \bibinfo{author}{Berkowitz, M.~L.}
\newblock \bibinfo{title}{Ewald summation for systems with slab geometry}.
\newblock \emph{\bibinfo{journal}{The Journal of Chemical Physics}}
  \textbf{\bibinfo{volume}{111}}, \bibinfo{pages}{3155--3162}
  (\bibinfo{year}{1999}).

\end{thebibliography}

\clearpage
\ifarXiv


\section*{Supplementary material}

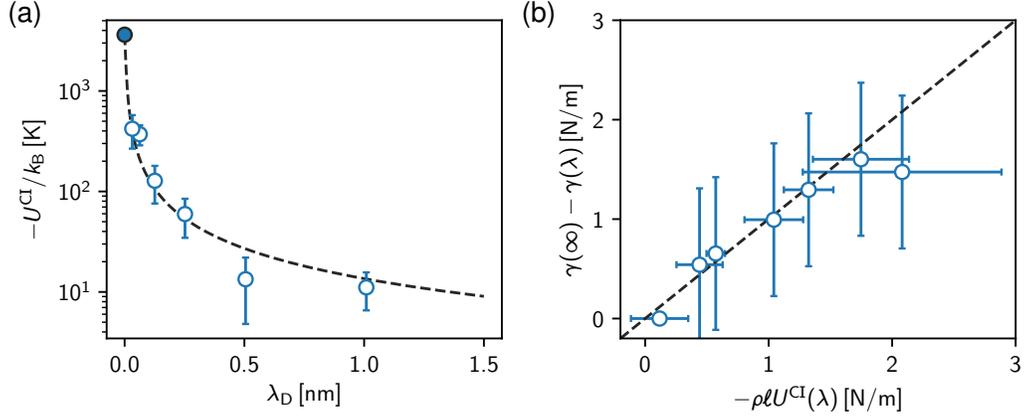

Figure S1. (a) Electrostatic energy per ion corresponding to the interactions between a confined charged liquid and induced charges in the confining metal $U^{\mathrm{CI}}$ as a function of their electrostatic screening length $\lambda_{\mathrm{D}}$. The dashed line, which is provided as a guide to the eye, corresponds to a fit $U^{\mathrm{CI}}(\lambda_{\mathrm{D}}) = U^{\mathrm{CI}}(0)[1 - \exp(-\lambda_0/\lambda_{\mathrm{D}})]$. As expected, $U^{\mathrm{CI}}(\lambda_{\mathrm{D}})$ converges to the analytically known solution for a perfect metal where the charge image contribution amounts to half the overall Coulomb interaction. Moreover, one can also verify that $U^{\mathrm{CI}}(\lambda_{\mathrm{D}}) \to 0$ for $\lambda_{\mathrm{D}} \to \infty$. (b) Scaling of the surface tension $\gamma(\lambda)$ at a metallic surface characterized by a screening parameter $\lambda$ with respect to its value at an insulating surface $\lambda = \infty$ as a function of the charge image interaction $\rho U^{\mathrm{CI}}(\lambda)$ (where $\rho$ is the ion density since $U^{\mathrm{CI}}(\lambda)$ is an energy per ion). The length $\ell$ converts a volume energy to a surface energy. The dashed line serves as a guide to the eye.

Table S1. Simulation parameters for the BMH potential employed for the salt–salt interaction[a].

|        | $A$ [kcal/mol] | $\sigma$ [Å] | $B$ [Å] | $C$ [kcal/mol/Å$^6$] | $D$ [kcal/mol/Å$^8$] |
|--------|----------------|--------------|---------|----------------------|----------------------|
| Na–Na  | 6.0811         | 2.340        | 0.317   | 24.1807              | 11.5146              |
| Cl–Cl  | 3.6487         | 3.170        | 0.317   | 1669.6786            | 3353.6227            |
| Na–Cl  | 4.8639         | 2.755        | 0.317   | 161.2044             | 200.0662             |

[a] Parameters taken from Ref. [1].

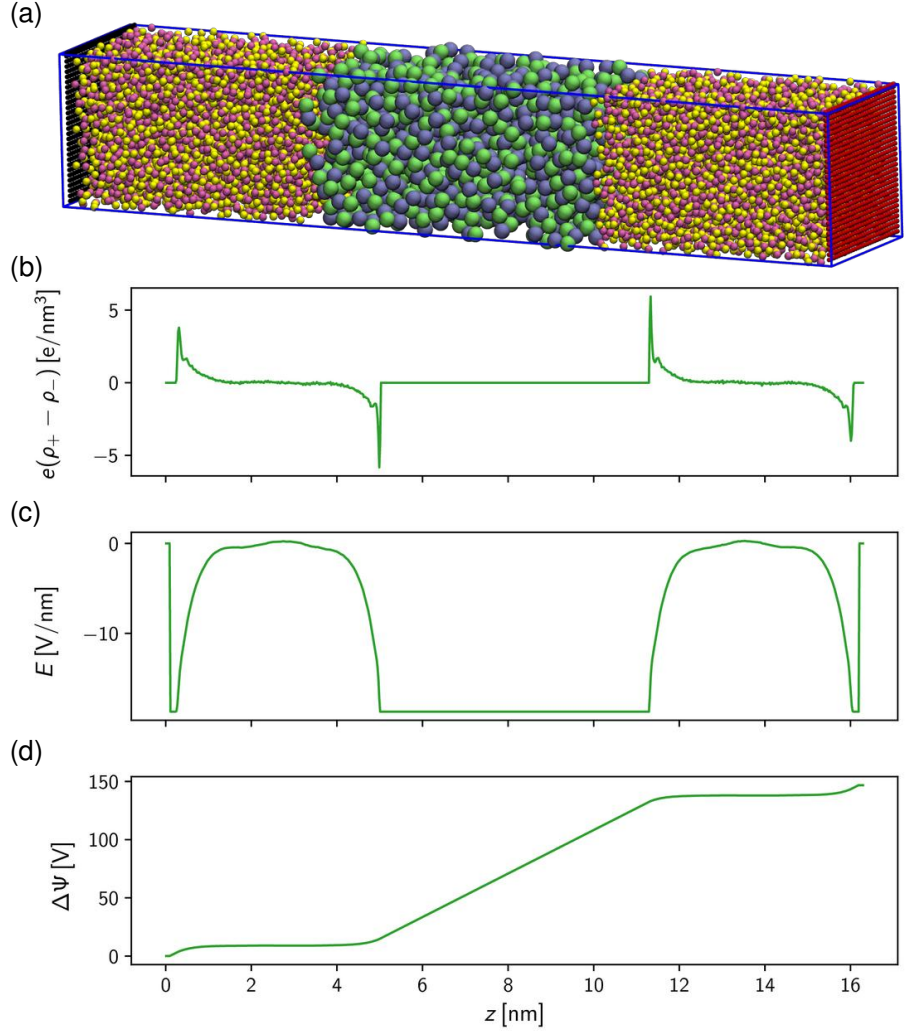

Figure S2. (a) Typical molecular configuration of the simulation set-up employed for the capacitance determination. The cathode (black) and anode (red) are made up of point charges arranged on a 1Å square grid. These charges are separated from the TF fluid (yellow and purple particles) using 1Å-thick reflective walls. The salt ions, which constitute the material sandwiched by the Thomas-Fermi fluid, are shown as green and blue spheres. (b) Charge density profile for a surface charge $Q/A = 0.167\,\mathrm{C/m^2}$ and a Thomas-Fermi length $\lambda = 0.5\,\mathrm{nm}$. For clarity, the delta peaks corresponding to the surface charge of the electrodes are not shown. (c) and (d) show the corresponding electric field $E(z)$ and electrostatic potential $\Psi(z)$ obtained by integrating twice the resulting charge density profile: $E(z) = \int_{-\infty}^{z} \mathrm{d}z'\, e(\rho_+ - \rho_-)/\varepsilon_0$ and $\Psi(z) = -\int_{-\infty}^{z} \mathrm{d}z'\, E(z')$, respectively.

4

## II. ELECTROSTATIC INTERACTIONS CLOSE TO A THOMAS–FERMI SUBSTRATE

### A. Thomas Fermi screening

In classical electrostatic theory, a point charge brought in front of a conducting solid is treated by considering an ideal metal where the induced electric field is perfectly screened in an infinitesimally small surface layer [2]. The corresponding interaction potential of the point charge with the metal can then be obtained using, as a purely mathematical tool, the method of image charges (i.e. the potential of an equal charge of opposite sign mirrored by the surface). At the molecular scale, this macroscopic description of perfect screening breaks down as the quantum mechanical nature of electrons leads to delocalization [3]. More in detail, the latter implies that the screening lengthscale cannot be infinitesimally small, but rather the interactions are screened over a typical lengthscale $\lambda$.

The simplest theory to describe screening at a finite wavevector $k_{\mathrm{TF}} = \lambda^{-1}$ was introduced independently by Thomas [4] and Fermi [5] who treated the electrons as a non-interacting homogeneous gas. The latter approximation neglects the correlations of electronic wavefunctions using a mean-field treatment. In this case, the chemical potential $\mu$ of the electrons equals (in the zero temperature limit) the Fermi energy $E_{\mathrm{F}}$ and is directly related to the electron density $n_0$ inside the metal (see e.g. Ref. [6] for a derivation),

$$\mu = \mathcal{E}_{\mathrm{F}} = \frac{\hbar^2}{2m_e} \left(3\pi^2 n_0\right)^{2/3}, \tag{S1}$$

where $\hbar$ is the reduced Planck constant and $m_e$ the electron mass.

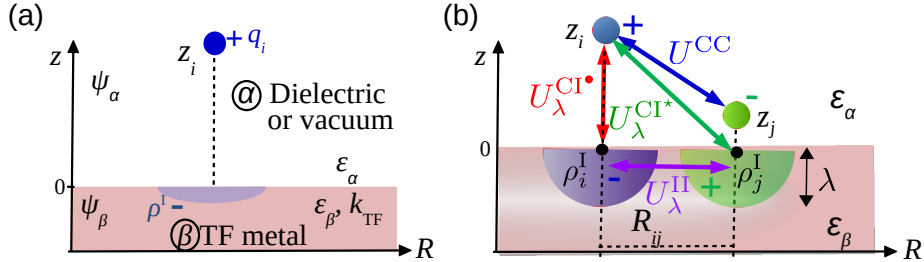

Figure S3. Electrostatic interactions at a metal/dielectric interface. (a) A single point charge $q_i$ in the dielectric medium $\alpha$ induces a charge distribution $\rho^{\mathrm{I}}$ inside the metal $\beta$ (but located close to the interface). (b) Decomposition of the total energy $U_\lambda$ for two charges $i$ and $j$ into one- and two-body terms.

Let us now consider an interfacial system consisting of a dielectric medium (denoted by $\alpha$) in contact with a metal (denoted by $\beta$), see Fig. S3(a). Any charge brought close to the metal will

disturb the local density distribution $n_0(\mathbf{r})$ by generating an electrostatic potential, $\Psi_\beta$, within the metal. The induced charge density inside the metal, $\rho^{\mathrm{I}}(\mathbf{r})$, results from a competition between (1) the energy reduction due to the screening of the electrostatic potential $\Psi_\beta$ generated by the external charge and (2) the energy cost of localizing the induced charge $\rho^{\mathrm{I}}$ [7],

$$\rho^{\mathrm{I}}(\mathbf{r}) = -e\left[n_0(\mu - e\Psi_\beta(\mathbf{r})) - n_0(\mu)\right]. \tag{S2}$$

Equation (S2) is the non-linear Thomas–Fermi (TF) equation and an expansion to the first order yields the linearized TF equation,

$$\rho^{\mathrm{I}}(\mathbf{r}) = -e^2 \frac{\partial n_0}{\partial \mu} \Psi_\beta(\mathbf{r}). \tag{S3}$$

As showN in Section VI, in analogy to the linear dielectric theory, it is convenient to define the proportionality factor as the Thomas–Fermi wavevector,

$$k_{\mathrm{TF}}^2 = \frac{e^2}{\varepsilon_0} \frac{\partial n_0}{\partial \mu}. \tag{S4}$$

For temperatures small compared to the Fermi temperature, i.e. $T \ll T_{\mathrm{F}} = \mathcal{E}_{\mathrm{F}}/k_{\mathrm{B}}$, $\partial n_0/\partial \mu$ is directly given by the density of states $\mathcal{D}(\mathcal{E}_{\mathrm{F}})$ at the Fermi level $\mathcal{E}_{\mathrm{F}}$ of the material $\beta$ [8]. Taking into account polarization effects via $\varepsilon_\beta$, the TF wavevector thus follows as

$$k_{\mathrm{TF}} = \lambda^{-1} = \frac{e^2}{\varepsilon_\beta \varepsilon_0} \mathcal{D}(\mathcal{E}_{\mathrm{F}}) = \frac{e^2}{\varepsilon_\beta \varepsilon_0} \frac{1}{2\pi^2} = \sqrt{m_e e^2 k_{\mathrm{F}}/(\varepsilon_\beta \varepsilon_0 \hbar^2 \pi^2)}, \tag{S5}$$

where $k_{\mathrm{F}} = \left(\mathcal{D}(\mathcal{E}_{\mathrm{F}})/(3\pi^2)\right)^{1/3}$ is the Fermi wave vector and

$$\mathcal{D}(\mathcal{E}_{\mathrm{F}}) = \frac{1}{2\pi^2} \left(\frac{2m}{\hbar^2}\right)^{3/2} \sqrt{\mathcal{E}_{\mathrm{F}}} \tag{S6}$$

is the density of states of a Fermi gas at the Fermi energy $\mathcal{E}_{\mathrm{F}} = \hbar^2 (3\pi^2 n_0)^{2/3}/(2m_e)$. Within the nearly free electron model, the charge density is roughly equal to the number density of atoms in the metal $n_0 \sim 10\text{-}100\,\mathrm{nm^{-3}}$. Thus, at room temperature, the relation $T \ll T_{\mathrm{F}} \sim 10^5\text{-}10^6\,\mathrm{K}$ is fully justified. $k_{\mathrm{TF}}$ is a quantity that characterizes the strength of the screening (see Section VI) which, according to Eq. (S5), increases with $\mathcal{D}(\mathcal{E}_{\mathrm{F}})$. An increased density of states permits the electron density to vary more with a lesser effect on the chemical potential of the local Fermi liquid. This reduces the cost of screening the external potential and, thus, results in a smaller screening length $\lambda$.

### B. Green function of a charge close to a Thomas–Fermi interface

The Green function of a charge $q_i$ at a distance $z_i$ from a TF interface allows obtaining the potential energy [Fig. S3(a)]. Due to the symmetry, we define $\Psi(\mathbf{r}) = \Psi(z, R)$ in cylindrical

coordinates. The derivation presented below largely follows the steps presented in Ref. [9]. The charge $q_i$ is located inside an insulator ($\varepsilon_\alpha \neq 1$) or vacuum ($\varepsilon_\alpha = 1$) at a position $\mathbf{r}_i$ with a coordinate $z_i$ along the $z$-direction normal to the surface and a radial position $R_i = 0$ in the $xy$-plane, while the TF substrate is located in the half-space $z < 0$, see Fig. S3(a).

The Green function for the upper half-space $z > 0$ is obtained from Poisson equation,

$$\nabla^2 \Psi_\alpha(\mathbf{r}) = \frac{\rho(\mathbf{r})}{\varepsilon_0 \varepsilon_\alpha} = -\frac{q_i \delta(\mathbf{r} - \mathbf{r}_i)}{\varepsilon_0 \varepsilon_\alpha} \tag{S7}$$

$$\rightarrow \nabla^2 \Psi_\alpha(z, R) = -\frac{q_i \delta(z - z_i)\delta(R)}{2\pi\varepsilon_0 \varepsilon_\alpha R}, \tag{S8}$$

where we have used the Dirac function in cylindrical coordinates, $\delta(\mathbf{r} - \mathbf{r}_i) = \delta(R - R_i)/2\pi R$. Inserting Eq. (S4) into Eq. (S3) and using Poisson equation given in Eq. (S7), the linearized TF equation can be recast as

$$\nabla^2 \Psi_\beta(z, R) - k_{\mathrm{TF}}^2 \Psi_\beta(z, R) = 0. \tag{S9}$$

Equations (S8) and (S9) describe the electrostatic potential $\Psi$ in the insulator $\alpha$ and the metal $\beta$, respectively. The Green function in Eqs. (S8) and (S9) can be solved for via Hankel integral transformation of first order [10]: $\Psi(z, K) = \int \mathrm{d}R R J_0(KR)\Psi(z, R)$ where $R$ is the radial component along the surface [see Fig. S3(a)], $K$ is the corresponding radial wavevector and $J_0$ is the Bessel function of first kind. Hankel transformation applied to Eqs. (S8) and (S9) yields

$$\left(\partial_{zz} - K^2\right) \Psi_\alpha = -\frac{q_1 \delta(z - z_i)}{2\pi\varepsilon_0 \varepsilon_\alpha} \tag{S10}$$

$$\left(\partial_{zz} - \kappa^2\right) \Psi_\beta = 0, \tag{S11}$$

where we have used $\kappa_{\mathrm{TF}}^2 = K^2 + k_{\mathrm{TF}}^2$ to simplify notation.

Using that the potential must vanish in all directions at infinity and the boundary condition at the surface given by the continuity of the potential $\Psi_\alpha(z_i; z = 0_+) = \Psi_\beta(z_i; z = 0_-)$ and the electric displacement field $\varepsilon_\alpha[\partial_z\Psi_\alpha](z_i; z = 0_+) = \varepsilon_\beta[\partial_z\Psi_\beta](z_i; z = 0_-)$, the Green functions read (as derived e.g. in Refs. [11–14])

$$\begin{aligned} \Psi_\alpha^>(z_i; z > z_i, K) &= \frac{q_1}{4\pi\varepsilon_0 \varepsilon_\alpha K} \left[\mathrm{e}^{+Kz_i} + \frac{\varepsilon_\alpha K - \varepsilon_\beta \kappa_{\mathrm{TF}}}{\varepsilon_\alpha K + \varepsilon_\beta \kappa_{\mathrm{TF}}}\mathrm{e}^{-Kz_i}\right] \mathrm{e}^{-Kz} \\ \Psi_\alpha^<(z_i; z < z_i, K) &= \frac{q_1}{4\pi\varepsilon_0 \varepsilon_\alpha K} \left[\mathrm{e}^{+Kz} + \frac{\varepsilon_\alpha K - \varepsilon_\beta \kappa_{\mathrm{TF}}}{\varepsilon_\alpha K + \varepsilon_\beta \kappa_{\mathrm{TF}}}\mathrm{e}^{-Kz}\right] \mathrm{e}^{-Kz_i} \end{aligned} \tag{S12}$$

$$\Psi_\beta(z_i; z, K) = \frac{q_1 \mathrm{e}^{-Kz_i}}{2\pi\varepsilon_0} \frac{\mathrm{e}^{\kappa_{\mathrm{TF}}z}}{\varepsilon_\alpha K + \varepsilon_\beta \kappa_{\mathrm{TF}}}. \tag{S13}$$

The first term inside the brackets of Eq. (S12) is the potential generated by the point charge $q_i$, whereas the second term corresponds to the potential generated by the induced charge in the TF substrate. We explicitly kept in Eqs. (S12) and (S13) the parametric dependence on the distance $z_i$.

### C. One-body interaction: A single point charge close to a Thomas–Fermi interface

Let us consider an isolated point charge $i$ at a distance $z_i$ from a TF metal, see Fig. S3(a). For an ideal metal $\lambda \to \infty$, the electrostatic energy can be obtained using the method of image charges as $U_{\infty}^{\text{CI} \bullet} = -q_i^2/(16\pi\varepsilon_0\varepsilon_\alpha z_i)$, which is equal to half of the energy that a real pair of interacting charges would have [15]. We recall that the symbol $\bullet$ refers to the interaction of a charge with its image in the metal, see main text. For a TF metal characterized by a finite screening length $\lambda$, we follow the derivation in Ref. [9] and compute the electrostatic energy via the volume integral of the product of the charge density and the potential, $U_\lambda^{\text{CI} \bullet} = \int d\mathbf{r} \rho(\mathbf{r}) \Psi(\mathbf{r})$. The latter integral has to be taken over the full space in $z$ and thus the corresponding solutions $\Psi_\alpha$ and $\Psi_\beta$ in Eqs. (S12) and (S13) have to been taken for $z > 0$ and $z < 0$, respectively. The total charge density $\rho(\mathbf{r}) = \rho^\bullet(\mathbf{r}) + \rho^\text{I}(\mathbf{r})$ follows from the sum of the point charge, $\rho^\bullet(\mathbf{r}) = q_i\delta(\mathbf{r} - \mathbf{r}_i)$ and the induced charge density $\rho^\text{I}(\mathbf{r})$.

By combining Eqs. (S3), (S4) and (S13), the induced charge density $\rho^\text{I}$ as derived from the TF equation is proportional to the electrostatic potential in the metal,

$$\rho^\text{I}(z_i; z, K) = -\varepsilon_0\varepsilon_\beta k_{\text{TF}}^2 \Psi_\beta(z_i; z, K) = -\frac{\varepsilon_\beta k_{\text{TF}}^2 q_1 e^{-Kz_i}}{2\pi\left(\varepsilon_\alpha K + \varepsilon_\beta \kappa_{\text{TF}}\right)} e^{\kappa_{\text{TF}} z}. \tag{S14}$$

Note that Eqs. (S13) and (S14) are defined only in the lower half-space $z < 0$. Upon applying the Plancherel theorem and the inverse Hankel transform $\Psi(z, R) = \int dK K J_0(KR) \Psi(z, K)$, the one-body energy reads [9]

$$
\begin{aligned}
U_\lambda^{\text{CI} \bullet}(z_i) &= \frac{1}{2} \int_{-\infty}^{\infty} dz \int_0^{\infty} 2\pi R dR \left[\rho^\text{I}(z_i; z, R)\Psi_\beta(z_i; z, R) + \rho^\bullet(z_i; z, R)\Psi_\alpha(z_i; z, R)\right] \\
&= \pi \int_{-\infty}^{\infty} dz \int_0^{\infty} dK K \left[\rho^\text{I}(z_i; z, K)\Psi_\beta(z_i; z, K) + \rho^\bullet(z_i; z, K)\Psi_\alpha(z_i; z, K)\right] \\
&\qquad\qquad =: U_\text{I}^{\text{CI} \bullet}(z_i) + U_\bullet^{\text{CI} \bullet}(z_i), \quad \text{(S15)}
\end{aligned}
$$

The first term defined by the right hand side of Eq. (S15) is the potential energy $U_\text{I}^{\text{CI} \bullet}$ of the induced charge, i.e. the energy needed to induce the charge density $\rho^\text{I}$ in the metal. The second term is the potential energy $U_\bullet^{\text{CI} \bullet}$ of the point charge in front of the Thomas–Fermi substrate. Inserting Eq. (S14) and using that the induced charge $\rho^\text{I}$ is only defined in the half-space $z < 0$,

one obtains

$$U_{\mathrm{I}}^{\mathrm{CI}\bullet}(z_i) = -\pi\varepsilon_0\varepsilon_\beta k_{\mathrm{TF}}^2 \int_{-\infty}^0 \mathrm{d}z \int_0^\infty \mathrm{d}K K \left[\Psi_\beta(z_i; z, K)\right]^2$$

$$= -\frac{q_i^2}{8\pi\varepsilon_0\varepsilon_\beta} k_{\mathrm{TF}}^2 \int_0^\infty \mathrm{d}K K \frac{\mathrm{e}^{-2Kz_i}}{\left((\varepsilon_\alpha/\varepsilon_\beta)K + \sqrt{K^2 + k_{\mathrm{TF}}^2}\right)^2 \sqrt{K^2 + k_{\mathrm{TF}}^2}}. \quad (S16)$$

By substituting the integration variable with the dimensionless variable $\xi = Kz_i$, the latter can be brought into a slightly more familiar form [9],

$$U_{\mathrm{I}}^{\mathrm{CI}\bullet}(z_i) = -\frac{q_{\mathrm{I}}^2}{16\pi\varepsilon_0\varepsilon_\beta} \int_0^\infty \mathrm{d}\xi \frac{2(k_{\mathrm{TF}}z_i)^2 \xi \mathrm{e}^{-2\xi}}{\left((\varepsilon_\alpha/\varepsilon_\beta)\xi + \sqrt{\xi^2 + (k_{\mathrm{TF}}z_i)^2}\right)^2 \sqrt{\xi^2 + (k_{\mathrm{TF}}z_i)^2}}$$

$$=: -\frac{q_{\mathrm{I}}^2}{16\pi\varepsilon_0\varepsilon_\beta} \mathcal{I}_{\mathrm{I}}(z_i, k_{\mathrm{TF}}), \quad (S17)$$

where the right hand side defines the integral $\mathcal{I}_{\mathrm{I}}$.

We now consider the second term in Eq. (S15). In cylindrical coordinates, the charge density located on the axis $R = 0$ reads as $\rho^\bullet(z) = q_i\delta(z - z_i)/(2\pi)$. Therefore, the two cases $\Psi_\alpha^>$ and $\Psi_\alpha^<$ in Eq. (S12) are equal upon integration in $z$ of $\rho^\bullet\Psi_\alpha$ in Eq. (S15). For instance, using the expression $\Psi_\alpha^>$ one obtains

$$U_\bullet^{\mathrm{CI}\bullet}(z_i) = \pi \int_{-\infty}^\infty \mathrm{d}z \int_0^\infty \mathrm{d}K K \frac{q_{\mathrm{I}}\delta(z - z_i)}{2\pi} \frac{q_{\mathrm{I}}}{4\pi\varepsilon_0\varepsilon_\alpha K} \left[\mathrm{e}^{+Kz_i} + \frac{\varepsilon_\alpha K - \varepsilon_\beta\kappa_{\mathrm{TF}}}{\varepsilon_\alpha K + \varepsilon_\beta\kappa_{\mathrm{TF}}}\mathrm{e}^{-Kz_i}\right]\mathrm{e}^{-Kz}$$

$$= \frac{q_{\mathrm{I}}^2}{8\pi\varepsilon_0\varepsilon_\alpha} \int_0^\infty \mathrm{d}K \left[1 - \frac{\varepsilon_\beta\sqrt{K^2 + k_{\mathrm{TF}}^2} - \varepsilon_\alpha K}{\varepsilon_\beta\sqrt{K^2 + k_{\mathrm{TF}}^2} + \varepsilon_\alpha K}\mathrm{e}^{-2Kz_i}\right] \quad (S18)$$

The first term corresponds to the diverging self-energy of any point charge. In line with the treatment in classical electrostatics [2], we omit this static part in the following, as it simply corresponds to a constant self-contribution. Again, substituting the integration variable by $\xi = Kz_i$ yields

$$U_\bullet^{\mathrm{CI}\bullet}(z_i) = \frac{q_{\mathrm{I}}^2}{8\pi\varepsilon_0\varepsilon_\alpha z_i} \int_0^\infty \mathrm{d}\xi \frac{\sqrt{\xi^2 + (k_{\mathrm{TF}}z_i)^2} - (\varepsilon_\alpha/\varepsilon_\beta)\xi}{\sqrt{\xi^2 + (k_{\mathrm{TF}}z_i)^2} + (\varepsilon_\alpha/\varepsilon_\beta)\xi}\mathrm{e}^{-2\xi}$$

$$= \frac{q_{\mathrm{I}}^2}{16\pi\varepsilon_0\varepsilon_\alpha z_i} \left[\int_0^\infty \mathrm{d}\xi \frac{4(\varepsilon_\alpha/\varepsilon_\beta)\xi}{\sqrt{\xi^2 + (k_{\mathrm{TF}}z_i)^2} + \varepsilon_\alpha/\varepsilon_\beta\xi}\mathrm{e}^{-2\xi} - 1\right]$$

$$=: -\frac{q_{\mathrm{I}}^2}{16\pi\varepsilon_0\varepsilon_\alpha z_i} \left[1 - \mathcal{I}_\bullet(z_i, k_{\mathrm{TF}})\right]. \quad (S19)$$

The integrals $\mathcal{I}_{\mathrm{I}}$ and $\mathcal{I}_\bullet$ defined above can be obtained numerically with high accuracy using common routines such as QUADPACK. In practice, we employ the latter through Python's `scipy.integrate.quad` interface to obtain the numerical resuslts such as those shown in Fig. S4(a).

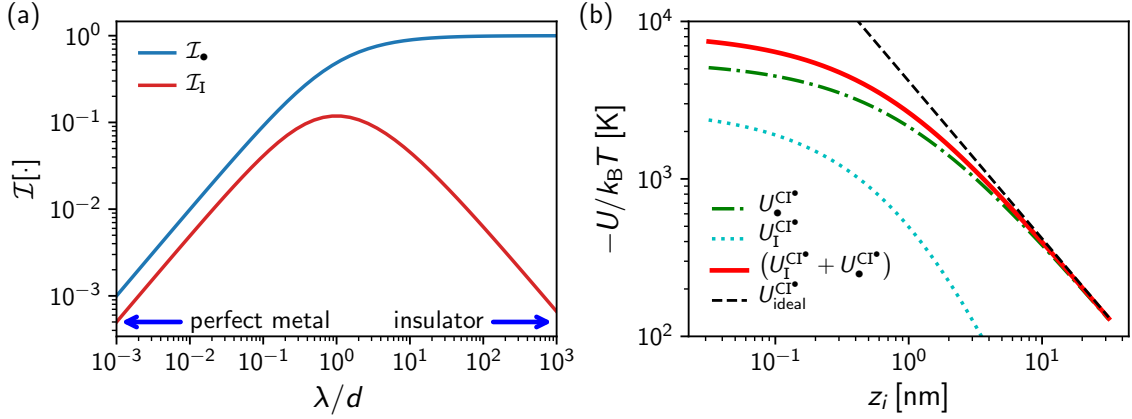

Figure S4. Numerical results for the one-body energy: (a) Dependence of the integrals $\mathcal{I}$ on the dimensionless screening length $\lambda/z_i = k_{\mathrm{TF}} z_i$. (b) One-body electrostatic energy $-U_\lambda^{\mathrm{CI}\bullet}$ and its contributions according to Eqs. (S17) and (S19) for a fixed screening length $\lambda = 1\,\mathrm{nm}$. The dashed black line shows the perfect metal limit.

Asymptotic analysis reveals that $\mathcal{I}_{\mathrm{I}}$ vanishes both in the ideal metal and insulator limits, $k_{\mathrm{TF}} = 1/\lambda \to \infty$ and $k_{\mathrm{TF}} \to 0$, respectively. $\mathcal{I}_{\bullet}$ vanishes in the ideal metal limit and goes to unity for the perfect insulator, $\varepsilon_\beta = \varepsilon_\alpha = 1$. This directly reveals that, without dielectric contrast, as expected, the energy of a single point charge at a perfect metal interface is half of the equivalent energy at an insulator surface. Figure S4(b) shows as an example the resulting energy $U^{\mathrm{CI}\bullet}$ and its contributions for $\lambda = 1\,\mathrm{nm}$ compared to the ideal metal, $U_\infty^{\mathrm{CI}\bullet} = q_1^2/(16\pi\varepsilon_0\varepsilon_\alpha z_i)$. As the implicit dielectric constants in our simulations are $\varepsilon_\beta = \varepsilon_\alpha = 1$, we fix this ratio for all data reported here. Note that for numerical reasons the unit of the inverse length $\xi = K z_i$ in Eqs. (S17) and (S19) should be chosen such that $\lambda/z_i \sim 1$

### D. Two-body interaction: Two point charges close to a Thomas–Fermi interface

Let us now consider two point charges $i$ and $j$ in front of a TF substrate as depicted in Fig. S3(b). As discussed in the main text, we decompose the total energy into the following contributions,

$$U_\lambda(z_i, z_j, R_{ij}) = U^{\mathrm{CC}}(r_{ij}) + U_\lambda^{\mathrm{CI}}(z_i, z_j, R_{ij}) + U_\lambda^{\mathrm{II}}(z_i, z_j, R_{ij}), \tag{S20}$$

where $z_i$ and $z_j$ are the distances normal to the surface, $R_{ij}$ is the in-plane separation and $r_{ij} = \sqrt{R_{ij}^2 + (z_i - z_j)^2}$ is the distance between the charges. In the following, we discuss the individual contributions and the numerical solution procedures employed.

**Direct Coulomb interaction**: $U^{\mathrm{CC}}$ denotes the direct Coulomb charge–charge interaction between the point charges, $U^{\mathrm{CC}} = q_i q_j / (4\pi\varepsilon_0 \varepsilon_\alpha r_{ij})$. In the general case of a fluid consisting of $N$ charges, the corresponding Coulomb energy of the charge $i$ reads as

$$U_i^{\mathrm{CC}} = \frac{q_i}{8\pi\varepsilon_0\varepsilon_\alpha} \sum_{j\neq i}^{N} \frac{q_j}{r_{ij}},$$
(S21)

where the additional factor $1/2$ stems from the fact that the energy $U^{\mathrm{CC}}$ contributes to both particles $i$ and $j$. To simplify notation, in the following, we skip the index $i$ for the total energy of particle $i$ and refer to the sum $j \neq i$ whenever no indices $(ij)$ are specified. This approach is equivalent to considering the statistical average of a particle in a fluid, $U^{\mathrm{CC}} = \langle U_i^{\mathrm{CC}} \rangle_N$.

A special case emerges when a periodic crystal of point charges $q_i = -q_j = q$ is considered. In this case, Eq. (S21) can be recast in terms of the Madelung constant $\mathcal{M}$,

$$U^{\mathrm{CC}} = -\frac{q^2}{8\pi\varepsilon_0\varepsilon_\alpha a_0} \mathcal{M}.$$
(S22)

Here, $a_0$ is the nearest neighbor distance (which corresponds to half of the crystal unit cell) and $\mathcal{M}$ depends on the dimension of the crystal as $\mathcal{M} = 2\ln 2$ in one dimension, $\mathcal{M} \approx 1.61554$ in two dimensions [16] and $\mathcal{M} \approx 1.747564594$ in three dimensions [17].

**Charge–image interaction $\mathbf{U}_\lambda^{\mathrm{CI}}$**: We decompose the charge–image interaction of particle $i$ interacting with particle $j$ at the TF interface according to $U_\lambda^{\mathrm{CI}} = U_\lambda^{\mathrm{CI}^\bullet}(z_i) + U_\lambda^{\mathrm{CI}^\circ}(z_i, z_j, R_{ij})$, where $U_\lambda^{\mathrm{CI}^\bullet}$ is the one-body term given by Eq. (S15) and $U_\lambda^{\mathrm{CI}^\circ}$ the electrostatic energy of charge $i$ interacting with the induced charge density $\rho^{\mathrm{I}}(z_j; z, R)$ due to the charge $j$. Making use of symmetry, the only relevant variable in the $R$-direction is the projected particle distance $R_{ij}$, see Fig. S3(b). $U_\lambda^{\mathrm{CI}^\circ}$ is obtained from the convolution of the induced charge of $j$ and the electrostatic potential $\Psi_\beta$ due to charge $i$ in the metal, i.e. $U_\lambda^{\mathrm{CI}^\circ} = \int d\mathbf{r} \Psi_\beta(\mathbf{r}_i; \mathbf{r}) \rho^{\mathrm{I}}(\mathbf{r}_j; \mathbf{r})$. This leads to

$$\begin{aligned}
U_\lambda^{\mathrm{CI}^\circ}(z_i, z_j, R_{ij}) &= \int_{-\infty}^{0} dz \int_0^\infty 2\pi R_{ij} dR_{ij} \Psi_\beta(z_i; z, R_{ij}) \rho^{\mathrm{I}}(z_j; z, R_{ij}) \\
&= 2\pi \int_{-\infty}^{0} dz \int_0^\infty dK K J_0(K R_{ij}) \Psi_\beta(z_i; z, K) \rho^{\mathrm{I}}(z_j; z, K).
\end{aligned}$$
(S23)

Using Eqs. (S13) and (S14), this yields

$$\begin{aligned}
U_\lambda^{\mathrm{CI}^\circ}(z_i, z_j, R_{ij}) &= -\frac{q_i q_j}{2\pi\varepsilon_0} \int_{-\infty}^{0} dz \int_0^\infty dK K J_0(K R_{ij}) \frac{\varepsilon_\beta k_{\mathrm{TF}}^2}{(\varepsilon_\alpha K + \varepsilon_\beta \kappa_{\mathrm{TF}})^2} e^{-K(z_i + z_j)} e^{2\kappa_{\mathrm{TF}} z} \\
&= -\frac{q_i q_j}{4\pi\varepsilon_0} \int_0^\infty dK J_0(K R_{ij}) \frac{K \varepsilon_\beta k_{\mathrm{TF}}^2}{\sqrt{K^2 + k_{\mathrm{TF}}^2} \left(\varepsilon_\alpha K + \varepsilon_\beta \sqrt{K^2 + k_{\mathrm{TF}}^2}\right)^2} e^{-K(z_i + z_j)}.
\end{aligned}$$
(S24)

Equation (S24) can be integrated numerically as discussed above.

**Image–image interaction $U_\lambda^{\mathrm{II}}$:** The resulting expressions for the image–image interactions are more involved. The general expression for the energy of the image charge of particle $i$ in the electrostatic potential induced by particle $j$ follows from the convolution

$$U_\lambda^{\mathrm{II}}(\mathbf{r}_i, \mathbf{r}_j) = \int \mathrm{d}\mathbf{r}\, \rho^{\mathrm{I}}(\mathbf{r}_i; \mathbf{r}) \Psi_\beta(\mathbf{r}_j; \mathbf{r}). \tag{S25}$$

Due to the spatial extension of the induced charges, the volume integral cannot be transformed into cylindrical coordinates conveniently. We thus use cartesian coordinates and, to simplify notation, we locate the charge $i$ at the origin, $\mathbf{r}_i = (0, 0, z_i)$ and the charge $j$ at distance $R_{ij}$ on the $x$-axis, $\mathbf{r}_j = (R_{ij}, 0, z_j)$,

$$U_\lambda^{\mathrm{II}}(z_i, z_j, R_{ij}) = \frac{1}{2} \int \mathrm{d}x \mathrm{d}y \mathrm{d}z\, \rho^{\mathrm{I}}(z_i; x, y, z) \Psi_\beta(z_j, R_{ij}; x, y, z). \tag{S26}$$

This allows us to use the previously derived expression for the induced charge given in Eq. (S14), which upon inverse Hankel transform reads

$$\rho^{\mathrm{I}}(z_i; x, y, z) = -\frac{\varepsilon_\beta k_{\mathrm{TF}}^2 q_i}{2\pi} \int_0^\infty \mathrm{d}K J_0(K r_\parallel) \frac{\mathrm{e}^{-K z_i}}{\varepsilon_\alpha K + \varepsilon_\beta \sqrt{K^2 + k_{\mathrm{TF}}^2}} \mathrm{e}^{\sqrt{K^2 + k_{\mathrm{TF}}^2} z}, \tag{S27}$$

where $r_\parallel = \sqrt{x^2 + y^2}$. Similarly, upon introducing $R_\parallel = \sqrt{(x - R_{ij})^2 + y^2}$, the electrostatic potential follows from Eq. (S13) as

$$\Psi_\beta(z_j, R_{ij}; x, y, z) = \frac{q_j}{2\pi\varepsilon_0} \int_0^\infty \mathrm{d}K J_0(K R_\parallel) \frac{\mathrm{e}^{-K z_j}}{\varepsilon_\alpha K + \varepsilon_\beta \sqrt{K^2 + k_{\mathrm{TF}}^2}} \mathrm{e}^{\sqrt{K^2 + k_{\mathrm{TF}}^2} z}. \tag{S28}$$

Performing the volume integral in Eq. (S26) is a daunting task. We thus employ a numerical integration in $\mathbf{r}$. Figure S5(a) shows the potential $\Psi_\beta$ for a point charge $q_i = 1\,\mathrm{e}$ located at a distance $z_i = 1\,\mathrm{nm}$ from a Thomas–Fermi metal characterized by a screening length $\lambda = 1\,\mathrm{nm}$. As expected, one observes that the potential and the corresponding induced charge density in Fig. S5(b) extend over a characteristic length $\sim 1\,\mathrm{nm}$. Contrary, for $\lambda = 0.05\,\mathrm{nm}$, the potential is screened on $\sim 0.05\,\mathrm{nm}$ in Fig. S5(c) and the induced charge density in Fig. S5(d) approaches a point-like distribution as expected for a good metal.

Having assessed the numerical solution of Eqs. (S27) and (S28), we obtain the electrostatic energy by calculating the product $u_\lambda^{\mathrm{II}} = \rho^{\mathrm{I}}(\mathbf{r}_i; \mathbf{r}) \Psi_\beta(\mathbf{r}_j; \mathbf{r})$ in Eq. (S26) on a mesh as shown in Fig. S6. Due to the approximately exponential decay, we employ a mesh of logarithmically increasing spacing with increasing distance to the position of the induced charges as indicated by the red lines in Fig. S6. We limit the integration to a distance from the charges of $10\lambda$ in the lateral direction

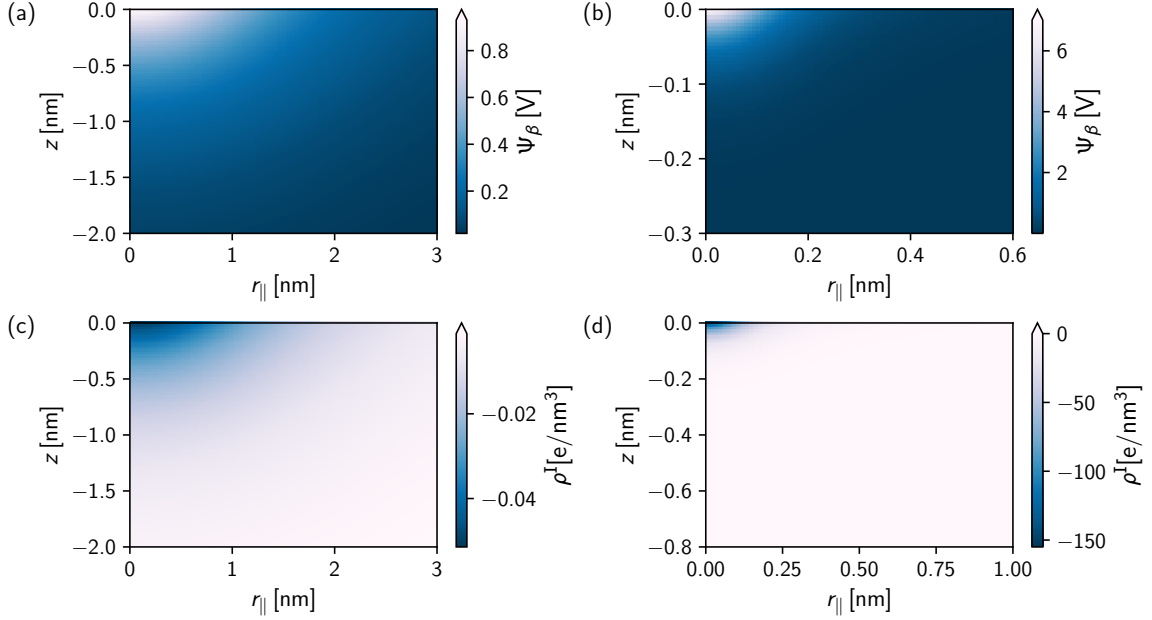

Figure S5. Electrostatic potential $\Psi_\beta$ and charge density $\rho^{\mathrm{I}}$ induced by a point charge $q_i = 1\mathrm{e}$ in front of a TF interface. (a) Electrostatic potential in the $y = 0$ plane for $z_i = 1\,\mathrm{nm}$ and $\lambda = 1\,\mathrm{nm}$. (c) shows the corresponding induced charge density $\rho^{\mathrm{I}}$. (b) and (d) show corresponding results for $z_i = 0.1\,\mathrm{nm}$ and $\lambda = 0.05\,\mathrm{nm}$.

and $8\lambda$ in the direction normal to the surface. The total image–image energy $U_\lambda^{\mathrm{II}} = \int \mathrm{d}\mathbf{r}\, u_\lambda^{\mathrm{II}}$ is then obtained by numerical integration using the corresponding volume element of each grid element. We explicitly checked that increasing the domain size does not alter the obtained energy as $u_\lambda^{\mathrm{II}} \sim 0$ at the boundary of the integration volume. Furthermore, due to the logarithmic grid employed, the integral converges well with the grid resolution as shown in Fig. S7. We characterize the meshing by a parameter $m$, which corresponds to the number of elements employed in the $z$-direction. Figure S7 reports the convergence of $\Delta U_m^{\mathrm{II}} = U_m^{\mathrm{II}}/U_{m=50}^{\mathrm{II}}$, where we consider $m = 50$ as a reference. In practice, we find that $m = 30$, corresponding to a number of grid elements $N_{\mathrm{grid}} \sim 10^5$, yields accurate results.

## III. THOMAS–FERMI ENERGY FOR A TWO-DIMENSIONAL IONIC CRYSTAL

Let us consider the total energy of a charge $i$ interacting with all other charges $j$ in the system, $U_\lambda^i = \sum_{i \neq j} U_\lambda^{ij}$. Taking the average over $N$ particles, $U_\lambda = \langle U_\lambda^i \rangle_N$, we obtain the total electrostatic energy divided by the number of particles. For a two-dimensional square ionic crystal, the energy

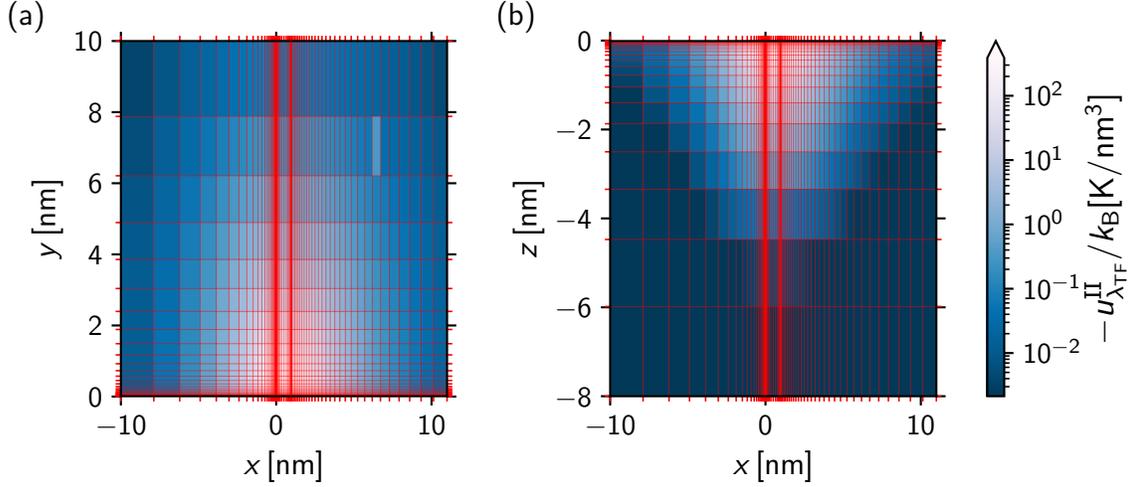

Figure S6. Energy density of the image–image interaction $u_\lambda^{\mathrm{II}}$ inside the TF substrate. The point charges $q_i = -q_j = 1\,\mathrm{e}$ are located at $\mathbf{r}_i = (0, 0, 1\,\mathrm{nm})$ and $\mathbf{r}_j = (1\,\mathrm{nm}, 0, 1\,\mathrm{nm})$, respectively. The TF substrate is characterized by a screening length $\lambda = 1\,\mathrm{nm}$. Results are shown for (a) the z=0 plane and (b) the $y = 0$ plane. Red lines indicate the logarithmic mesh employed.

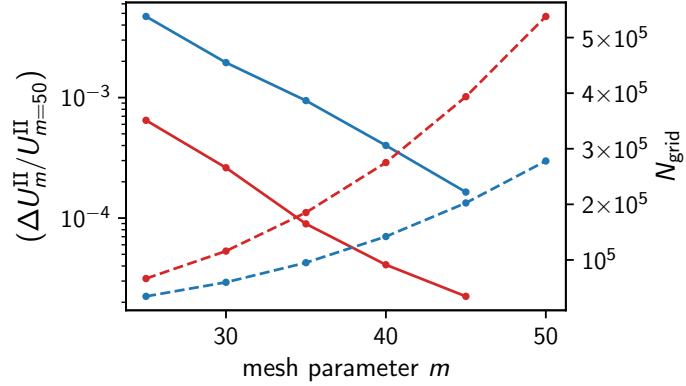

Figure S7. Convergence of the numerical integration of the image–image interaction energy with the resolution of the spatial grid (see Fig. S6). The mesh parameter $m$ corresponds to the number of elements in the $z$-direction. The corresponding total number of elements $N_{\mathrm{grid}}$ is shown as dashed lines (right axis). For illustration, data are shown for two point charges at $z_i = z_j = 1\,\mathrm{nm}$, $R_{ij} = 2\,\mathrm{nm}$ using $\lambda = 1\,\mathrm{nm}$ (blue lines) and $z_i = z_j = 0.1\,\mathrm{nm}$, $R_{ij} = 2\,\mathrm{nm}$, $R_{ij} = 2\,\mathrm{nm}$ using $\lambda = 0.05\,\mathrm{nm}$ (red lines).

only depends on the distance $d$ from the TF substrate and the crystal lattice spacing $a_0$. The expression corresponding to Eq. (S20) is thus given by

$$U_\lambda = U^{\mathrm{CC}}(a_0) + U^{\mathrm{CI}^\bullet}(d) + U_\lambda^{\mathrm{CI}^\circ}(d, a_0) + U_\lambda^{\mathrm{II}}(d, a_0).  \tag{S29}$$

The direct Coulomb interaction, which is independent of the TF substrate, only depends on the crystal lattice spacing $a_0$. $U^{\mathrm{CC}}(a_0)$ is thus conveniently expressed in terms of a Madelung constant as discussed in Section II D. The one-body contribution $U^{\mathrm{CI}^\bullet}(d)$, which only depends on $d$, is discussed in Section II C. Contrary, $U_\lambda^{\mathrm{CI}^\circ}$ and $U_\lambda^{\mathrm{II}}$ explicitly depend on the electrostatic screening characterized by $\lambda$ and, thus, cannot be expressed in terms of a general Madelung constant. We explicitly perform the sum over all neighbors in the $x, y$-plane,

$$U_\lambda^{\mathrm{CI}^\circ/\mathrm{II}}(d, a_0) = \sum_{\substack{j,k=-\infty \\ k \neq j}}^{\infty} U_\lambda^{\mathrm{CI}^\circ/\mathrm{II}} \left( z_i = d, z_2 = d, R_{ij} = a_0 \sqrt{j^2 + k^2} \right) \Big|_{q_j = (-1)^{j+k}}.  \tag{S30}$$

In practice, we cut the infinite sum in Eq. (S30) at a finite number of neighbors $N_{\mathrm{neigh}}$. We find that $N_{\mathrm{neigh}} \sim 10$ yields reasonable accuracy which we monitor by fitting the energy of the form $U(N_{\mathrm{neigh}}) = U(\infty) + U^\circ / N_{\mathrm{neigh}}$. Figure S8 shows $U_\lambda^{\mathrm{CI}^\circ}(N_{\mathrm{neigh}})$ for $\lambda = 0.75\,\mathrm{nm}$, $d = 0.8\,\mathrm{nm}$ and $a_0 = 1.475\,\mathrm{nm}$ as employed in Fig. 3 of the main text. We perform a bootstrapping analysis discarding one data point each time during the fit to obtain the family of fitting parameters shown as lines in Fig. S8(a) and from which we obtain confidence intervals for $N_{\mathrm{neigh}} \to \infty$. To obtain $U_\lambda^{\mathrm{CI}^\circ/\mathrm{II}}(d, a_0)$, we interpolate the values obtained at a set of distances $d$ using a weighted smoothing spline as shown in Fig. S8(b). The extrapolated numerical error estimate shown in the inset of Fig. S8(b) is, in all cases, negligible compared to the total charge–induced energy $U^{\mathrm{CI}}$ shown in Fig. 3(a) of the main text.

## IV. INFLUENCE OF INTERACTION POTENTIALS, SALT SLAB WIDTH AND THOMAS–FERMI LAYER WIDTH

To justify our choice of parameters $n = 8$, $d_{\mathrm{w}} = 10\,\mathrm{nm}$ and $d_{\mathrm{TF}} = 20\,\mathrm{nm}$, we here fix $\lambda^\star = 0.5\,\mathrm{nm}$ and explicitly vary these parameters. Figure S9 shows the data from Fig. 3 in the main text (blue circles) together with the numerical solution of the TF model (solid black line). When varying the repulsive power law exponent to $n = 12$, the resulting energies (green triangles in Fig. S9) are in perfect agreement with $n = 8$. Upon increasing the distance between the two Thomas–Fermi interfaces to $d_{\mathrm{w}} = 40\,\mathrm{nm}$ (red squares), the energies close to the surface (small $d$) agree well. However, for $d \gtrsim 0.2\,\mathrm{nm}$ the energy is slightly smaller and in better agreement with the TF

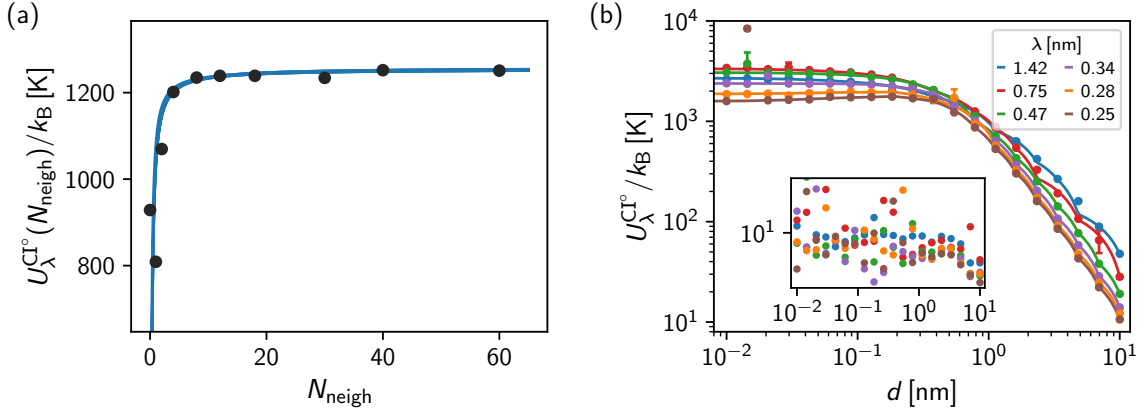

Figure S8. (a) Extrapolation of the neighbor sum for $U^{CI°}$ as defined by Eq. (S30). Fits of the form $U(\infty) + U°/N_{neigh}$ are performed leaving out one data point each and shown as lines. (b) The resulting values $U^{CI°}(N \to \infty)$ at varying distance for given $\lambda$ are interpolated using a weighted smoothing spline. The inset shows the estimated numerical error from the fits in (a) which, in practice, is negligible. The same procedure is employed to obtain $U^{II}$.

model prediction for a single interface (shown as solid black line in Fig. S9), hinting to possible interactions with the second interface.

In Fig. S10 we assess robustness of our results with respect to the influence of different TF layer thickness $d_{TF}$. Similar to $d_w$, an increase of $d_{TF}$ enhances the agreement of $U_\lambda$ with the TF model in Fig. S10(a). For our analysis provided in the main text, $d_{TF} = 10\,nm$ was used (shown as red squares in Fig. S10(a)). The energy of the TF fluid $U_\lambda^{TF}$ converges exponentially with $d_{TF}$ with a decay length of about $5\,nm$, see Fig. S10 (b), where we show $\Delta u_\lambda^{TF} = u_\lambda^{TF}(d_{TF}) - u_\lambda^{TF}(\infty)$ normalized by the extrapolated value at infinite layer thickness $d_{TF}$. To conclude, increasing $d_w$ and $d_{TF}$ slightly increases the agreement with the TF model of a single interface at the cost of significantly increased simulation effort (due to both an increase of the number of particles $[\sim d_{TF}]$ and the number of grid points necessary for the same accuracy using the PPPM method $[\sim d_{TF} + d_w]$). Contrary, it is important to note that our simulation approach can deal with a finite thickness of the metallic substrate $d_{TF}$.

## V. ENERGY DECOMPOSITION FROM SIMULATION RERUNS

As discussed in the main text, the simulated electrostatic energy $\tilde{U}_\lambda(d)$ consists of all ion pair contributions in Eq. (S20) and a contribution $U_\lambda^{TF}$ corresponding to the self Thomas–Fermi fluid

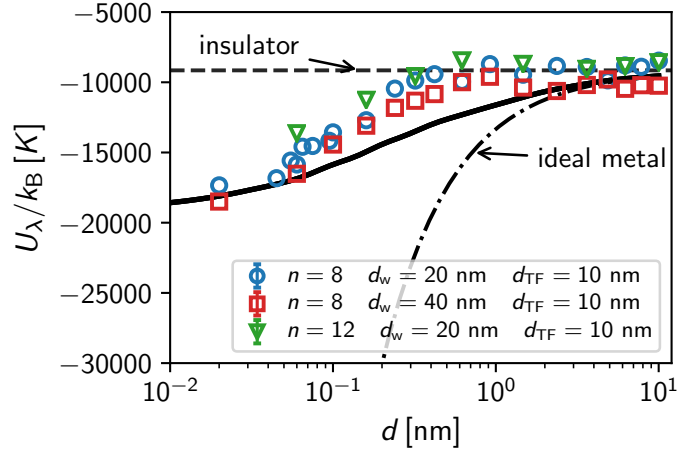

Figure S9. Influence of the TF–TF interaction potential (via the power $n$) and wall–wall separation $d_\mathrm{w}$ on the resulting energy $U_\lambda$. All values are shown for $\lambda = 0.47\,\mathrm{nm}$ (see main text). The solid line denotes the numerical solution of the TF model.

energy in the absence of the confined system. The total electrostatic energy between two charges $i$ and $j$ as measured in the molecular simulations thus reads:

$$\tilde{U}_\lambda(r_{ij}) = \tilde{U}^\mathrm{CC}(z_i, z_j, R_{ij}) + \tilde{U}^\mathrm{CI}_\lambda(z_i, z_j, R_{ij}) + \tilde{U}^\mathrm{II}_\lambda(z_i, z_j, R_{ij}) + \tilde{U}^\mathrm{TF}_\lambda(d_\mathrm{TF}). \tag{S31}$$

The latter contribution $\tilde{U}^\mathrm{TF}_\lambda$ can be seen as the ground-level energy of the Thomas–Fermi fluid in the simulation; the interaction between induced charges $\tilde{U}^\mathrm{II}_\lambda$ correspond accordingly to the change in its energy with respect to this reference. To compare the simulation/theoretical energies, $\tilde{U}^\mathrm{TF}_\lambda$ must be removed from $\tilde{U}_\lambda(d)$ obtained in the simulation (since $\tilde{U}^\mathrm{TF}_\lambda$ is set to zero by definition in the Thomas–Fermi theory). The individual terms can be accessed from reruns of the simulation trajectory. In detail, using configurations obtained for $q = 1$ and $q_\mathrm{TF} \neq 0$, we re-evaluate the electrostatic energy with either the salt charge $q$ or the charge of the TF fluid $q_\mathrm{TF}$ set to zero. To compensate for the absence of screening between the periodic images, two-dimensional boundary conditions [18] are employed. The resulting energies read as

$$\tilde{U}_\lambda(r_{ij})\Big|_{q_\mathrm{TF}=0} = \tilde{U}^\mathrm{CC}(z_i, z_j, R_{ij}) \qquad \text{and} \tag{S32}$$

$$\tilde{U}_\lambda(r_{ij})\Big|_{q=0} = \tilde{U}^\mathrm{II}_\lambda(z_i, z_j, R_{ij}) + \tilde{U}^\mathrm{TF}_\lambda(d_\mathrm{TF}). \tag{S33}$$

Subtracting Eqs. (S32) and (S33) from Eq. (S31) yields the contribution $\tilde{U}^\mathrm{CI}_\lambda$

$$\tilde{U}_\lambda(r_{ij}) - \tilde{U}_\lambda(r_{ij})\Big|_{q_\mathrm{TF}=0} - \tilde{U}_\lambda(r_{ij})\Big|_{q=0} = \tilde{U}^\mathrm{CI}_\lambda(z_i, z_j, R_{ij}). \tag{S34}$$

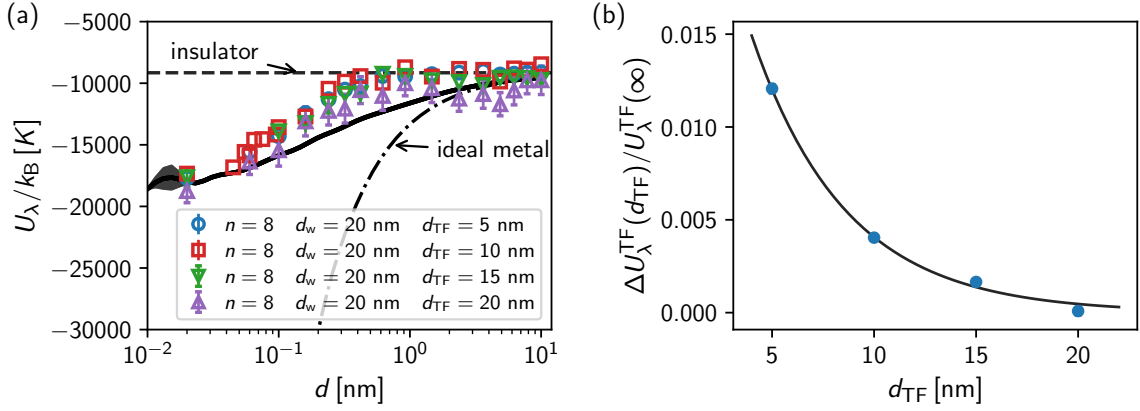

Figure S10. (a) Influence of $d_{TF}$ on the electrostatic energy $U_\lambda$ at fixed effective Thomas–Fermi length $\lambda = 0.47\,\mathrm{nm}$ (see main text). $n = 8$ and $d_w = 20\,\mathrm{nm}$ correspond to the values used for Fig. 3 of the main text, where $d_{TF} = 10\,\mathrm{nm}$ was used. The solid line denotes the numerical solution of the TF model as discussed in the main text. (b) The disjoining energy of the Thomas–Fermi fluid decays exponentially with a decay length of $\sim 5\,\mathrm{nm}$ (solid line).

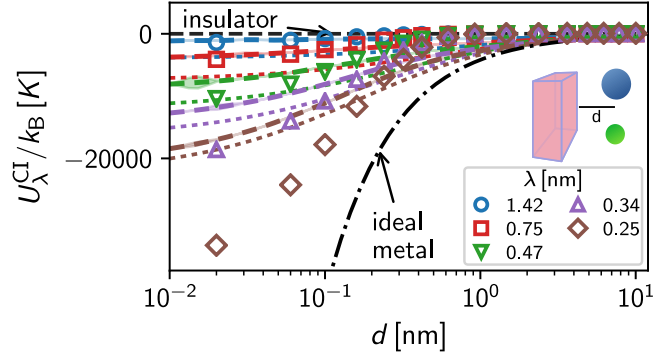

Figure S11. Energy $U_\lambda^{CI}(d)$ between a 2D ionic crystal and a Thomas–Fermi metal separated by a distance $d$ for different $\lambda$. For each $\lambda$, the symbols correspond to the effective simulation while the dashed line shows the Thomas–Fermi model (the dotted line is the one-body contribution $U_\lambda^{CI\bullet}$ to $U_\lambda^{CI}$). The black dash-dotted line shows the energy at a perfect metal surface $U_\lambda^{CI}(d) = U_\lambda^{CI\bullet}(d) \sim 1/d$. The black dashed line $U_\lambda^{CI}(d) = 0$ corresponds to data for an insulating surface.

Fig. S11 shows $U_\lambda^{CI}$ as a function of $d$ for different $\lambda$. $U_\lambda^{CI}$ decays with $d$ and, more importantly, varies between the values for an insulator [$U_\lambda^{CI}(d) = 0 \; \forall d$] and a perfect metal [$U_\lambda^{CI}(d) = U_\lambda^{CI\bullet}(d) = e^2/(16\pi\varepsilon_\alpha\varepsilon_0 d)$, i.e. the charge image model].

To decompose $\tilde{U}_\lambda^{II}$ and $\tilde{U}_\lambda^{TF}$, we perform a separate set of simulations with the TF fluid confined

between two reflecting walls and using the slab correction of Ref. [18] to mimic an insulating vacuum.

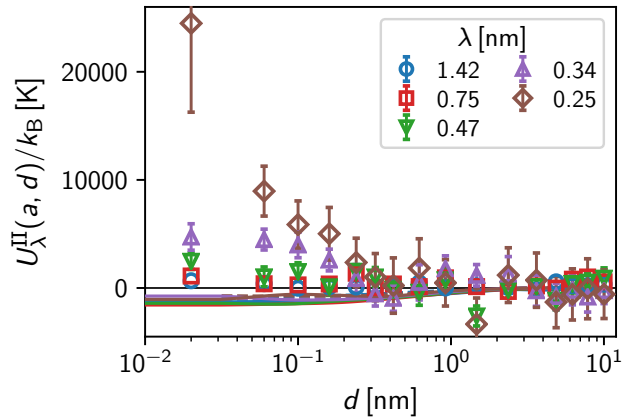

Figure S12. Electrostatic energy due to the induced charge densities, $U_\lambda^{\mathrm{II}}$, for the systems discussed in Fig. 2 of the main text. Data points show the simulated values obtained according to Eq. (S33), lines show the results from numerical integration of Eq. (S25) and extrapolating the neighbor sum as described in Section III.

In Figure S12, we show $U^{\mathrm{II}}$ obtained from Eq. (S25) and using the extrapolation of the neighbor sum described in Section III. In general, $U^{\mathrm{II}}$ is small compared to $U^{\mathrm{CI}}$ shown in Fig. 3(a) of the main text. Upon decreasing $\lambda$ (better metal), the induced charge density is more localized and the potential gets screened on shorter distances. Consequently, this term decays to zero in the perfect metal limit. As discussed in the main text, our simulations (data in Fig. S12) show the opposite behavior due to the imperfect nature of the TF fluid. The configuration of lowest energy for a system of point charges is the homogeneous distribution and localizing any charge distribution to form $\rho^{\mathrm{I}}$ necessarily increases this energy. This differs from the ideal electron gas behavior in the TF model, which has zero energy in the homogeneous case, where by inducing a charge density the resulting energy will become negative (lines in Fig. S12). Note that this deviation from ideality is captured in the rescaling of the effective screening $\tilde{\lambda}$ introduced in the main text.

## VI. GENERAL CONNECTION TO SCREENING IN LINEAR DIELECTRIC MEDIA

To obtain an explicit expression for the screened potential $\Psi_\beta$, we follow the basic equations of classical theory of electric polarization [2]. We consider a perturbing charge distribution $\rho^{\mathrm{ext}}(\mathbf{r})$

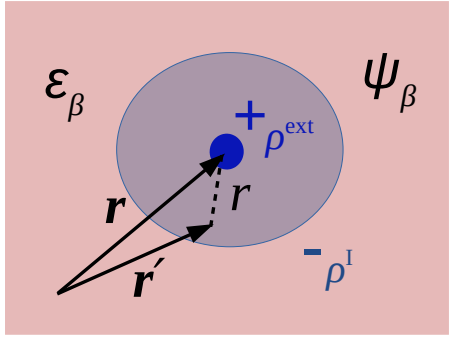

Figure S13. Illustration of a perturbing charge distribution $\rho^{\text{ext}}$ and the induced charge density $\rho^{\text{I}}$ inside the metal $\beta$, respectively.

located inside the metal, see Fig. S13. According to Poisson equation, $\rho^{\text{ext}}$ creates a potential $-\nabla^2 \Psi^{\text{ext}}(\mathbf{r}) = \rho^{\text{ext}}(\mathbf{r})/\varepsilon_0 \varepsilon_\beta$, which in turn induces a charge density $\rho^{\text{I}}$. Denoting the full charge density $\rho(\mathbf{r}) = \rho^{\text{ext}}(\mathbf{r}) + \rho^{\text{I}}(\mathbf{r})$, the full physical potential is $-\nabla^2 \Psi(\mathbf{r}) = \rho(\mathbf{r})/\varepsilon_0 \varepsilon_\beta$. In analogy to linear dielectric media, a linear relation between $\Psi$ and $\Psi^{\text{ext}}$ yields

$$\Psi^{\text{ext}}(\mathbf{r}) = \int \mathrm{d}\mathbf{r}' \varepsilon(\mathbf{r}, \mathbf{r}') \Psi(\mathbf{r}')., \tag{S35}$$

where we have introduced the non-local dielectric response function $\varepsilon(\mathbf{r}, \mathbf{r}')$. For a spatially uniform electron gas translational invariance can be used, i.e. $\varepsilon(\mathbf{r}, \mathbf{r}') = \varepsilon(|\mathbf{r} - \mathbf{r}'|)$. Equation (S35) can solved for $\Psi_\beta$ using Fourier transform and the convolution theorem [19]

$$\Psi(\mathbf{k}) = \frac{1}{\varepsilon(\mathbf{k})} \Psi^{\text{ext}}(\mathbf{k}). \tag{S36}$$

Equation (S36) shows that, for each wavevector $\mathbf{k}$, the total electrostatic potential $\Psi$ is given by the external potential $\Psi^{\text{ext}}$ at the same $\mathbf{k}$ but screened by a factor $1/\varepsilon(\mathbf{k})$, the Thomas–Fermi dielectric constant [7].

We now use a linear relation between the (Fourier transformed) induced charge density and the electrostatic potential, $\rho^{\text{I}}(\mathbf{k}) = -\chi(\mathbf{k})\Psi(\mathbf{k})$, where $\chi(\mathbf{k})$ is the dielectric susceptibility and $\varepsilon(\mathbf{k}) = (1 + \chi(\mathbf{k}))\varepsilon_0$. Comparison with Eq. (S3) yields the TF dielectric susceptibility [7],

$$\chi_{\text{TF}}(\mathbf{k}) = e^2 \frac{\partial n_0}{\partial \mu}. \tag{S37}$$

Thus, it is convenient to define the TF wavevector as given in Eq. (S4)

$$k_{\text{TF}}^2 = \frac{e^2}{\varepsilon_0} \frac{\partial n_0}{\partial \mu}, \tag{S38}$$

from which the Thomas–Fermi dielectric constant follows as

$$\varepsilon(\mathbf{k}) = 1 + \frac{k_{\mathrm{TF}}^2}{k^2}. \tag{S39}$$

To shed light on the significance of $\lambda$, we now consider the case of a point charge $\rho^{\mathrm{ext}}(\mathbf{r}) = q\delta(\mathbf{r} - \mathbf{r}')$ as defined in Fig. S3(b). Using $\mathbf{r} - \mathbf{r}' = r$, the external potential of the point charge and its Fourier transform directly follow from Poisson equation as

$$\Psi^{\mathrm{ext}}(r) = \frac{q}{4\pi\varepsilon_0 r} \qquad \text{and} \qquad \Psi^{\mathrm{ext}}(k) = \frac{q}{\varepsilon_0 k^2}. \tag{S40}$$

Using Eqs. (S36) and (S39), the total potential in the metal is given by

$$\Psi(k) = \frac{1}{\varepsilon(k)} \Psi^{\mathrm{ext}}(k) = \frac{q}{\varepsilon_0 \left(k^2 + k_{\mathrm{TF}}^2\right)}, \tag{S41}$$

which, upon inverse Fourier transformation, yields

$$\Psi(r) = \int \frac{\mathrm{d}k}{(2\pi)^3} \mathrm{e}^{ik\mathbf{r}} \frac{q}{\varepsilon_0 \left(k^2 + k_{\mathrm{TF}}^2\right)} = \frac{q}{4\pi\varepsilon_0 r} \mathrm{e}^{-k_{\mathrm{TF}} r}. \tag{S42}$$

Equation (S42) has the form of a screened Coulomb potential and $\lambda = k_{\mathrm{TF}}^{-1}$ is analogous to the Debye screening length $\lambda_{\mathrm{D}}$ obtained for electrolyte solutions [20]. This analogy forms the motivation for the explicit TF fluid screening approach in the present work.

---

[1] J. Anwar, D. Frenkel, and M. G. Noro, J. Chem. Phys. **118**, 728 (2002).

[2] J. D. Jackson, *Classical Electrodynamics* (Wiley, New York, 1999).

[3] N. D. Lang and W. Kohn, Phys. Rev. B **7**, 3541 (1973).

[4] L. H. Thomas, Math. Proc. Camb. Philos. Soc. **23**, 542 (1927).

[5] E. Fermi, Rend Accad Naz Lincei **6**, 32 (1927).

[6] R. Baer, "Electron density functional theory," (2016).

[7] N. W. Ashcroft and N. D. Mermin, *Solid State Physics* (Holt, Rinehart and Winston, 1976).

[8] The electrons obey the Fermi–Dirac distribution function,

$$n_0(\mu) = \mathcal{D}(\mu) \times \frac{1}{\exp\left(\frac{\mu - \mathcal{E}_{\mathrm{F}}}{k_{\mathrm{B}} T} + 1\right)},$$

such that in the limit $\mathcal{E}_{\mathrm{F}}/k_{\mathrm{B}}T \to \infty$ the second term results in a Heaviside step function, $n_0(\mu) = \mathcal{D}(\mu)\Theta(\mu - \mathcal{E}_{\mathrm{F}})$. Taking the derivative $\partial n_0/\partial \mu = \mathcal{D}(\mu)\delta(\mu - \mathcal{E}_{\mathrm{F}}) = \mathcal{D}(\mathcal{E}_{\mathrm{F}})$ yields the density of states valid for $T \ll T_{\mathrm{F}}$.

[9] V. Kaiser, J. Comtet, A. Niguès, A. Siria, B. Coasne, and L. Bocquet, Faraday Discuss. **199**, 129 (2017).

[10] I. Bronshtein, K. Semendyayev, G. Musiol, and H. Mühlig, *Handbook of Mathematics : With 132 Tables*, 6th ed. (Springer, 2015).

[11] D. M. Newns, J. Chem. Phys. **50**, 4572 (1969).

[12] J. C. Inkson, J. Phys. C: Solid State Phys. **6**, 1350 (1973).

[13] A. A. Kornyshev, A. I. Rubinshtein, and M. A. Vorotyntsev, Phys. Status Solidi B **84**, 125 (1977).

[14] R. R. Netz, Phys. Rev. E **60**, 3174 (1999).

[15] M. M. Taddei, T. N. C. Mendes, and C. Farina, Eur. J. Phys. **30**, 965 (2009).

[16] J. Lekner, Physica A: Statistical Mechanics and its Applications **176**, 485 (1991).

[17] S. Tyagi, Prog Theor Phys **114**, 517 (2005).

[18] I.-C. Yeh and M. L. Berkowitz, J. Chem. Phys. **111**, 3155 (1999).

[19] Using the following definition of the Fourier transform of a function $f$:

$$f(\mathbf{k}) = \int \mathrm{d}r e^{-i\mathbf{k}\mathbf{r}} f(\mathbf{r}) \qquad \text{and} \qquad f(\mathbf{r}) = \frac{\mathrm{d}\mathbf{k}}{(2\pi)^3} e^{i\mathbf{k}\mathbf{r}} f(\mathbf{k})$$

.

[20] P. Debye and E. Hückel, Phys Zft **24**, 305 (1923).


\fi

\end{document}